\documentclass[useAMS,referee,usenatbib,usegraphicx]{biom}
\usepackage{algorithm,amsfonts,amsmath,amssymb,dsfont,placeins,xr-hyper}
\usepackage[colorlinks,allcolors=NavyBlue]{hyperref}
\usepackage[dvipsnames]{xcolor}
\graphicspath{{figures/}}

\makeatletter
\renewcommand*{\fps@algorithm}{tbp}
\makeatother



\title[The Multivariate Bernoulli detector]{The Multivariate Bernoulli detector:\\ Change point estimation in discrete survival analysis}

\author{Willem van den Boom$^{1,*}$\email{vandenboom@nus.edu.sg}, 
Maria De Iorio$^{1,2,**}$\email{mdi@nus.edu.sg},
Fang Qian$^{1,***}$\email{qianxiaoxie@gmail.com}, and 
Alessandra Guglielmi$^{3,****}$\email{alessandra.guglielmi@polimi.it} \\
$^{1}$Yong Loo Lin School of Medicine, National University of Singapore \\
$^{2}$Singapore Institute for Clinical Sciences, Agency for Science, Technology and Research \\
$^{3}$Department of Mathematics, Politecnico di Milano, Milan, Italy}

\begin{document}

%  This will produce the submission and review information that appears
%  right after the reference section.  Of course, it will be unknown when
%  you submit your paper, so you can either leave this out or put in 
%  sample dates (these will have no effect on the fate of your paper in the
%  review process!)

% \date{{\it Received October} 2007. {\it Revised February} 2008.  {\it
% Accepted March} 2008.}

%  These options will count the number of pages and provide volume
%  and date information in the upper left hand corner of the top of the 
%  first page as in published papers.  The \pagerange command will only
%  work if you place the command \label{firstpage} near the beginning
%  of the document and \label{lastpage} at the end of the document, as we
%  have done in this template.

%  Again, putting a volume number and date is for your own amusement and
%  has no bearing on what actually happens to your paper!  

% \pagerange{\pageref{firstpage}--\pageref{lastpage}} 
% \volume{64}
% \pubyear{2008}
% \artmonth{December}

%  The \doi command is where the DOI for your paper would be placed should it
%  be published.  Again, if you make one up and stick it here, it means 
%  nothing!

% \doi{10.1111/j.1541-0420.2005.00454.x}

%  This label and the label ``lastpage'' are used by the \pagerange
%  command above to give the page range for the article.  You may have 
%  to process the document twice to get this to match up with what you 
%  expect.  When using the referee option, this will not count the pages
%  with tables and figures.  

\label{firstpage}

%  put the summary for your paper here

\begin{abstract}
Time-to-event data are often recorded on a discrete scale with multiple, competing risks as potential causes for the event. In this context, application of continuous survival analysis methods with a single risk suffer from biased estimation. Therefore, we propose the Multivariate Bernoulli detector for competing risks with discrete times involving a multivariate change point model on the cause-specific baseline hazards. Through the prior on the number of change points and their location, we impose dependence between change points across risks, as well as allowing for data-driven learning of their number. Then, conditionally on these change points, a Multivariate Bernoulli prior is used to infer which risks are involved. Focus of posterior inference is cause-specific hazard rates and dependence across risks. Such dependence is often present due to subject-specific changes across time that affect all risks. Full posterior inference is performed through a tailored local-global Markov chain Monte Carlo (MCMC) algorithm, which exploits a data augmentation trick and MCMC updates from non-conjugate Bayesian nonparametric methods. We illustrate our model in simulations and on ICU data, comparing its performance with existing approaches.
\end{abstract}

%  Please place your key words in alphabetical order, separated
%  by semicolons, with the first letter of the first word capitalized,
%  and a period at the end of the list.
%


%  As usual, the \maketitle command creates the title and author/affiliations
%  display 

\maketitle

%  If you are using the referee option, a new page, numbered page 1, will
%  start after the summary and keywords.  The page numbers thus count the
%  number of pages of your manuscript in the preferred submission style.
%  Remember, ``Normally, regular papers exceeding 25 pages and Reader Reaction 
%  papers exceeding 12 pages in (the preferred style) will be returned to 
%  the authors without review. The page limit includes acknowledgements, 
%  references, and appendices, but not tables and figures. The page count does 
%  not include the title page and abstract. A maximum of six (6) tables or 
%  figures combined is often required.''

%  You may now place the substance of your manuscript here.  Please use
%  the \section, \subsection, etc commands as described in the user guide.
%  Please use \label and \ref commands to cross-reference sections, equations,
%  tables, figures, etc.
%
%  Please DO NOT attempt to reformat the style of equation numbering!
%  For that matter, please do not attempt to redefine anything!

\section{Introduction}
\label{sec:intro}

Time-to-event data are common in many applications such as finance, medicine and engineering with examples including time to payment, and survival and failure times.
Most approaches to survival data consider continuous event times. Nevertheless, it is more and more common to record time-to-event data on a discrete scale (e.g.\ patient-reported
outcomes, or time to pregnancy measured in number of menstrual cycles it takes a couple to conceive). See \citet{Schmid2020} for further examples. Discrete recording of the timings of events \citep{Allison1982} may 
occur when time is truly discrete, or when continuous time is partitioned
into non-overlapping intervals (corresponding, for instance, to days, weeks, or months)
and only the
interval in which an event occurs is recorded \citep{King2021}. This special case of interval censoring is usually referred to as \emph{grouped time} and, in this work, we consider discrete survival models that
arise as probabilistic grouped versions of  continuous-time frailty models from survival analysis
(see, e.g.,
\citeauthor{Hougaard1986}, \citeyear{Hougaard1986}, \citeyear{Hougaard1995};
\citeauthor{Andersen1993}, \citeyear{Andersen1993}, Chapter~IX;
\citeauthor*{Hougaard1994}, \citeyear{Hougaard1994};
\citeauthor{Kalbfleisch2002}, \citeyear{Kalbfleisch2002}).
Indeed, direct application of continuous-time methods to discretely recorded data may result in biased estimation \citep{Lee2018}.

Moreover, our focus is on discrete survival data in the presence of
multiple, competing risks that can cause an event. Traditional analyses often consider a single risk with events due to other risks, e.g.\ other causes of death, treated as censoring.
However, this generally violates the common assumption of
independent censoring \citep{Schmid2020}.
Moreover, such analyses can lead to misestimation of hazards and covariate effects
\citep{Andersen2012}.
Here we consider data from the
Medical Information Mart for Intensive Care IV \citep[\mbox{MIMIC\nobreakdash-IV},][]{Johnson2023} database
on length of stay in an intensive care unit (ICU), typically reported
discretely in days. While the MIMIC\nobreakdash-IV database documents admission and discharge times down to the minute, it is recommended to perform survival analysis using discrete day units. This approach is preferred because admission and discharge times within a single day are significantly determined by hospital protocols and staff decisions, rather than purely reflecting the patients' health conditions. The study considers three competing events that can terminate an ICU stay: discharge to home (69.0\%), transfer to another
medical facility (21.4\%), and in-hospital death (6.1\%).

The two main approaches for competing risks with discrete times are cause-specific hazard functions and the subdistribution hazard model \citep{Schmid2020}.
The latter is more suitable when interest is in one out of many risks.
The first approach usually exploits methods from generalized linear models (GLMs), enabling maximum likelihood estimation, variable selection and other methods from GLMs such as in
\citet{Tutz1995} and \citet*{Most2016}.
Our work places itself within this approach, focussing on scenarios with few risks and thus considers cause-specific hazards, i.e.\ a hazard function for each risk, introducing dependence across risks by building on recent advances in change point analysis.

A characteristic of traditional discrete survival models as compared to GLMs is that they present
unconstrained baseline hazards. This leads to a  large number of parameters to estimate and, as a consequence, to unstable estimation, especially if for certain time points the number of events is small.
To improve stability, regularizations of hazard functions have been proposed. See, for example,
\citet*{Luo2016}, \citet{Heyard2019} and \citet{Most2016}.
\citet{Fahrmeir1996} and \citet{King2021} employ random walks to smooth the hazard function.
In all these works, cause-specific baseline hazards are treated independently.
\citet{Vallejos2017} focus on risk-specific covariate selection, still assuming independence across risks. 
On the other hand, dependence across risks is plausible since multiple hazards can be affected by changes to the individual across time, and as such should be incorporated in the  model.

The main methodological contribution of this work
lies in modeling explicit and interpretable dependencies across risks. We introduce a multivariate change point model for baseline hazards, which offers two key advantages \citep{Kozumi2000}. Firstly, it reduces the number of parameters, promoting parsimony. Secondly, it accommodates the variability in hazard rates across different survival times. We adopt a Bayesian approach, assigning priors on the number and location of overall change points, inducing marginal dependence among them. For each overall change point, we use a Multivariate Bernoulli prior to determine which risks are involved, a method previously applied in time series analysis (\citeauthor*{Dobigeon2007}, \citeyear{Dobigeon2007}; \citeauthor{Harle2016}, \citeyear{Harle2016}). We term our approach the Multivariate Bernoulli detector, building on prior work by \citet{Harle2016}.

Change points have been widely studied in continuous survival analysis \citep*[e.g.][]{Matthews1982,Goodman2011}
but less so in the discrete case:
\citet{Kozumi2000} considers a single risk modeled via a Markov chain with a prespecified number of change points, while
\citet{Wang2007} use posterior predictive checks for change point detection, without allowing for covariates in the model.
On the other hand, we allow for covariates in the model and perform  cause-specific variable selection. Moreover, we estimate the number of change points,
testing for the presence and location of change points using Bayes factors or posterior probabilities.

The paper is structured as follows.
Section~\ref{sec:model} introduces the model and describes a tailored Markov chain Monte Carlo (MCMC) sampler.
Section~\ref{sec:application} discusses an application to the ICU data.
Section~\ref{sec:compare} compares our approach with existing ones. We conclude the paper in 
Section~\ref{sec:discussion}.

\section{Model}
\label{sec:model}

\subsection{Setup and notation}
\label{sec:notation}

We follow the notation in \cite{tutz2016modeling}.
Let $T_i$
denote the time to event for individual $i\in\{1,\dots,n\}$
where $n$ is the number of individuals.
In the discrete-time setting,
$T_i\in\{1,\dots,t_{\max}+1\}$
for some maximum time $t_{\max}$. The random variable
$T_i$ can for instance arise as the discretization, also known as grouping \citep{Kalbfleisch2002}, of a latent continuous time $T_i^c$ into $t_{\max}+1$ intervals
$[\omega_0,\omega_1),\,[\omega_1,\omega_2),\,\dots,\,[\omega_{t_{\max}}, \infty)$ with $\omega_0=0$. In this case there is a one-to-one correspondence between the set of integers $\{1,\dots,t_{\max}+1\}$ and the intervals of the real line where the continuous-time random variables $T^c_i$ are defined.
As such, the interpretation of $T_i=t_{\max} +1$ is similar to censoring, in the sense  that  the event will occur at time $t > t_{\max}$
(i.e.\ $T_i^c \geq \omega_{t_{\max}}$) 
and $t_{\max}+1$ is  simply a ``latent time'' that groups together individuals for which it is known that the even has not occurred up to time $t_{\max}$.
The time-to-event distribution is usually characterized by the overall hazard function
$\lambda(t\mid \theta_i) = {P(T_i=t\mid T_i\geq t,\theta_i)} = {P\{T_i^c\in [\omega_{t-1},\omega_t) \mid T_i^c\geq \omega_{t-1},\theta_i\}}$
for some vector of parameters $\theta_i$.

Additionally, we assume that observations are subject to censoring.
That means
that only a portion of the observed times can be considered as exact survival times. Let $C_i$ be the censoring time of individual $i$.
$C_i$ assumes values in
$\{1,\dots,t_{\max}\}$,
with $T_i$ and $C_i$ independent (random censoring). Moreover, we assume that the censoring mechanism is non-informative, i.e.\ it does not
depend on any parameters used to model the event times \citep{Schmid2020}.
Let
$\delta_i = \mathds{1}[T_i\leq C_i]$
be a censoring indicator,
where $\mathds{1}[\cdot]$ denotes the indicator function,
and $t_i=\min(T_i, C_i)$
the observed time.

We consider competing risks with $m$ different types of events.
For instance, the events can correspond to death due to $m$ different causes.
Then, $R_i\in\{1,\dots,m\}$ denotes the event type experienced by individual $i$ at time $T_i$,
for which we observe a value $r_i$ only in the absence of censoring, i.e.\ $\delta_i=1$.
Finally,
the cause-specific hazard function
is ${\lambda_r(t\mid \theta_i)} = {P(T_i=t,\, R_i = r\mid T_i\geq t,\theta_i)}$
such that
$\lambda(t\mid \theta_i) = \sum_{r=1}^m {\lambda_r(t\mid \theta_i)}$.

\subsection{Likelihood}
\label{sec:likelihood}

We assume independence across individuals such that the likelihood is a product over individual-specific terms.
For ease of explanation, we consider the likelihood contribution of one individual and drop subscripts $i$ in the remainder of this section unless otherwise specified.
Under the assumption that $C$ is independent of $T$ and $\theta$,
the likelihood for $\theta$ is given by 
\begin{equation} \label{eq:likelihood}
    P(T=t,\, R=r\mid \theta)^{\delta}\, P(T>t\mid \theta)^{1-\delta} \\
    = 
    \lambda_{r}(t\mid\theta)^{\delta}\,
    \prod_{l=1}^{t - \delta} \{1 - \lambda(l\mid\theta)\}
\end{equation}
and we specify $\lambda_{r}(t\mid\theta)$
and thus
$\lambda(t\mid\theta)$ according to the multinomial logit model which is the most popular for categorical responses \citep{tutz2016modeling}.
Specifically,
${\lambda_{r}(t\mid\theta)} = \exp(\eta_{rt}) / \{1 + \sum_{\rho=1}^m \exp(\eta_{\rho t})\}$
where $\eta_{rt} = \alpha_{rt}+\bm x^\top\bm\beta_r$ is a cause- and time-specific linear predictor, and $\theta = \{\eta_{rt}\}$.
The intercept $\alpha_{rt}$ represents the cause-specific baseline hazard.
The $p$-dimensional vector
$\bm\beta_r\ =(\beta_{r1},\ldots, \beta_{rp})$ consists of
the cause-specific regression coefficients of the covariates
in the $p$-dimensional vector
$\bm x\ =(x_1,\ldots,x_p)$. Such likelihood specification is also known as  proportional continuation ratio model \citep{tutz2016modeling} as increasing $x_j$ by one unit increases the cause-specific odds by the factor
$\exp(\beta_{rj})$. Note that, when $r=1$ and $\delta=1$, $\lambda_r(t\mid \theta)$ corresponds to $\phi_t$ defined in Section~\ref{sec:tvgeom} and the considerations on prior specification there discussed will be relevant when building a change point model on $\alpha_{rt}$.

\subsection{The Multivariate Bernoulli detector}
\label{sec:mvbd}

We propose a model on the baseline hazards that is flexible, yet has interpretable structure.
Specifically,
the sequence $\alpha_{r1},\ldots,\alpha_{r,t_{\max}}$
is set to follow a piecewise constant function.
We do so through a change point model with dependence across risks $r$.
In our setup, a change point corresponds to a time point where the hazard of at least one risk $r$ changes.
We specify a hierarchical prior on the change points which has three main components: (i) a prior on the number of change points; (ii) a prior on the location of change points; (iii) and a prior on which risks (at least one) have a change point at that particular time location, given that a change point at time~$t$ occurs.

\subsubsection{Prior specification on overall change points}
\label{sec:loc_cp}

In this section, we describe prior specification on the number and location of change points.
Let $\bm \alpha_t = (\alpha_{1t},\dots,\alpha_{mt})$.
Then, $\gamma_t = \mathds{1}[\bm\alpha_t\ne \bm\alpha_{t-1}]$ indicates whether there is an overall change point at $t\in\{1,\dots,t_{\max}\}$, i.e.\ if the hazard of at least a risk changes at time $t$.
Furthermore,
$K=\sum_{t\in\mathcal{T}}\gamma_t$
denotes the number of change points.
Here $\mathcal{T}$ defines the set of possible change point locations.
We specify the prior on $\bm\gamma = (\gamma_1,\dots,\gamma_{t_{\max}})$
hierarchically, by specifying a prior $p(K)$ on the number of change points and then $p(\bm\gamma\mid K)$
as this provides explicit regularization on $K$:
i.i.d.\ $\gamma_t$ would imply a binomial distribution on $K$.

To motivate the next model development,
consider the bottom plots in Figure~\ref{fig:tvgeom}
where two observations with $t_i=7$ out of 500 are not used when inferring change points.
Then, the lack of observations at time $t=7$ results in spurious change points at that time location and the next.
We restrict our inference to avoid such sensitivity to a few observations:
to aid identifiability, considering the flexibility of the underlying Time-Varying Geometric distribution which is discussed in Section~\ref{sec:tvgeom},
we only allow change points for a subset of times $\mathcal{T} \subset\{1,\dots,t_{\max}\}$ such that $\gamma_t = 0$ if $t\notin \mathcal{T}$.
Firstly, as it is typical in change point applications, we do not allow change points at the support boundary, in our case $t=1$
and $t=t_{\max}$. Moreover,
we do not allow a change point at time $t$
if both $t$ and $t-1$ have no observed events as the data lack information on which of the two time points would be a change point.
Also, we do not allow change points at a time $t$ with no observed events
if both neighboring times $t-1$ and $t+1$ have observed events, because this would lead to spurious change points due to the flexibility of the underlying Time-Varying Geometric, as seen in Figure~\ref{fig:tvgeom} (bottom row).
On the other hand, we prefer to introduce parsimony in the estimation of change points to improve interpretability.
We explore the effect of the restriction on change point locations in a simulation study in Web Section~\ref{sec:simul_restrict}. There, the restriction (i)~does not deteriorate inference, even if the true change point is not in $\mathcal{T}$; (ii)~avoids spurious change points at time locations without observed events.

We assume a Geometric distribution with success probability $\pi_K$ truncated to $K \leq |\mathcal{T}|$ as prior on the number of change points. We denote such prior as
$\mathrm{Geo}_{|\mathcal{T}|}(\pi_K)$.
For the locations of overall change points given $K$,
we use the uniform distribution on possible configurations $p(\bm\gamma\mid K) = 1 / \binom{|\mathcal{T}|}{K}$. In summary, the joint prior on the number and location of change points has a hierarchical specification: 
$p(K, \bm\gamma)= p(K)\, p(\bm\gamma \mid K)$.

\subsubsection{Cause-specific change point configuration}

In this section,  we discuss the prior on which risks present a jump in the hazard, given the vector  $\bm \gamma$.
Let
$z_{rt} = \mathds{1}[\alpha_{rt} \ne \alpha_{r(t-1)}]$
be an indicator variable denoting if a change point occurs at time $t$
for risk $r$.
If there is no change point at $t$ for any $r$, then $\gamma_t=0$
and
$z_{rt}=0$.
If $\gamma_t=1$,
then $z_{rt} =1$ for at least one $r$.

Let  
$\bm z_t = (z_{1t},\dots, z_{mt})$. 
Conditionally on $\gamma_t = 1$, we assume that $\bm z_t$ follows a
Multivariate Bernoulli distribution \citep[e.g.][]{Teugels1990}.
An $m$-dimensional binary vector $\bm z_t$ can assume $2^m$ possible values corresponding to a combination of $z_{rt}\in \{0,1\}$. The Multivariate Bernoulli distribution is then parameterized by a $2^m$-dimensional vector, whose elements correspond to the probability of each particular outcome (i.e.\ configuration). In our case, when modeling $\bm z_t$ given $\gamma_t =1$, we exclude the configuration of all zeros, i.e.\ $z_{rt} =0$ for every $r$. Thus, $\bm z_t$ can assume only $2^m -1$ possible values. We denote such distribution as $\mathrm{Ber}_0(\bm \psi)$, where $\bm \psi$ denotes the $(2^m -1)$- dimensional vector of configuration probabilities. In summary, the prior specification for $\bm z_t $ is
\[
\bm z_t \mid \gamma_t \sim \left\{
 \begin{array}{lcl}
\mathrm{Ber}_0(\bm \psi) &\quad & \mbox{if } \gamma_t =1 \\
\delta_{\bm 0 } &\quad &  \mbox{if } \gamma_t =0
   \end{array}
\right. 
\]
where $\delta_{\bm 0 }$ is a point mass at the zero vector.
We refer to the joint prior on $(K,\bm \gamma, \bm z)$ as Multivariate Bernoulli detector, where $\bm z= \{ \bm z_t \}_{t=1}^{t_{\max}}$.

\subsection{Further prior specification}
\label{sec:prior}

Model specification is completed by specifying independent prior distributions on
$\alpha_{rt}$ and $\bm\beta_r$. We specify a prior on $\alpha_{rt}$ conditionally on the number and location of change points. Since $\bm \alpha_r = (\alpha_{r1}, \ldots,\alpha_{rt_{\max}})$ is a piecewise constant function for each risk $r$, assuming constant values between change points, let $\alpha_{r\ell}^\star$ denote the unique value of $\alpha_{rt}$ over each time interval for risk $r$. Note that for each risk a change point can be activated or not, with the only constraint that a change point needs to be activated for at least one risk.
We assume
$\alpha_{r\ell}^\star
\sim \mathcal{N}(\mu_\alpha, \sigma^2_\alpha)$
independently across $\ell$ and $r$.

Furthermore, to identify important effects,
we assume a variable selection prior for the regression coefficients,
$\bm\beta_r = (\beta_{r1},\dots,\beta_{rp})$, which allows for risk-specific variable selection. Here, we consider a spike-and-slab prior \citep{Mitchell1988}:
$\beta_{rj}\sim \pi_\beta\, \mathcal{N}(0, \sigma^2_\beta) + (1-\pi_\beta)\, \delta_0$
where
$\pi_\beta$ is the prior inclusion probability.
We use the hyperprior $\pi_\beta \sim \mathcal{U}(0, 1)$.
In the application in Section~\ref{sec:application},
some variables are grouped as they are dummy variables associated with
a categorical covariate.
We modify the prior accordingly to perform groupwise variable selection as detailed in Web Appendix~\ref{ap:var_sel}.
We note that other possible prior choices are available in the literature to perform variable selection, such as shrinkage priors \citep{Bhadra2019}, which offer computational advantages, at the cost of depending on arbitrary thresholds to identify relevant effects.

In the simulation studies and the application on ICU data, we set the parameters as follows:
$\sigma^2_\beta = 1$, $\pi_K = 0.5$,
all elements of
$\bm \psi$ equal to $1 / (2^m - 1)$,
$\mu_\alpha = -9$ and $\sigma^2_\alpha = 3$.
The particular prior choice for $\alpha_{r\ell}^\star$ derives from the interpretation of the model in terms of the Time-Varying Geometric. In Section~\ref{sec:tvgeom}, 
we highlight the importance of shrinking the probabilities $\phi_t$ towards zero. This is equivalent, in absence of covariates, to  
shrinking $\exp(\alpha^\star_{r\ell})/ \{1+ \exp(\alpha^\star_{r\ell})\}$ towards zero and, consequently, $\alpha^\star_{r\ell}$ towards $-9$. Roughly speaking, a $\mathcal{N}(-9,3)$ on $\alpha^\star_{r\ell}$ is equivalent to a $\mathrm{Beta}(0.01, 1)$ prior on $\phi_t$, which is shown to have a good performance in Section~\ref{sec:tvgeom}.
Finally, we note that we could specify a prior on $\bm\psi$ to favor sparsity or a large number of change points.

\subsection{Rationale behind modeling strategy}
\label{sec:tvgeom}

\begin{figure}
\centering
\includegraphics{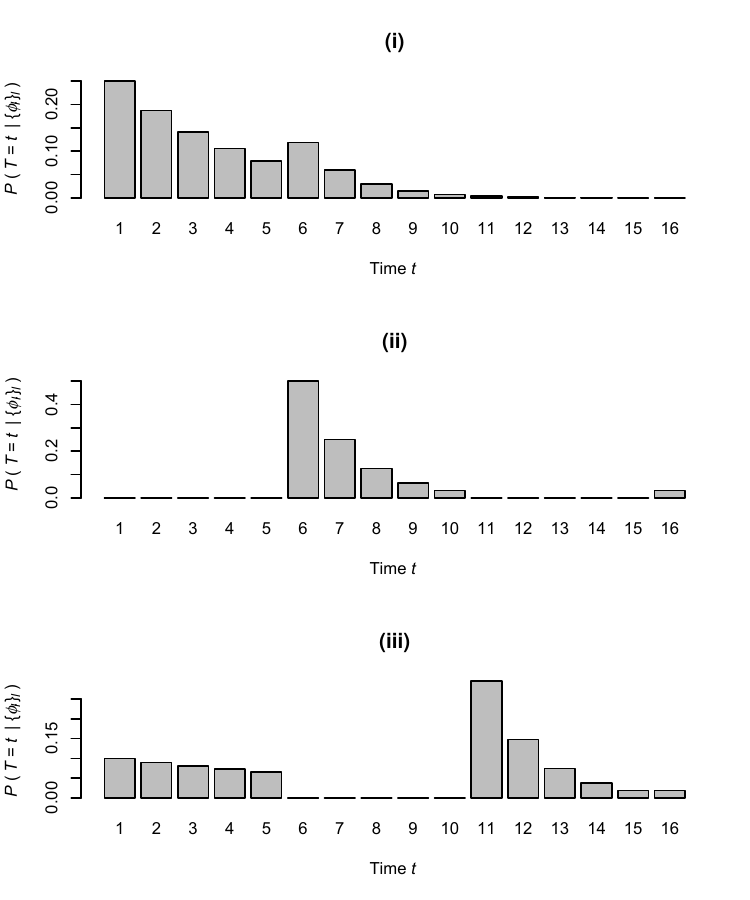}
\caption{Probability mass function of the Time-Varying Geometric distribution: Visualizations of \eqref{eq:tvgeom} with $t_{\max}=15$ and success probabilities (i) $\phi_t = 0.25$ for $t\leq 5$ and $\phi_t = 0.5$ otherwise; (ii) $\phi_t = 0$ for $t\leq 5$ or $t\geq 11$, and $\phi_t = 0.5$ otherwise; (iii) $\phi = 0.1$ for $t\leq 5$, $\phi_t = 0.5$ for $t\geq 11$ and $\phi_t = 0$ otherwise.}
\label{fig:tvgeom_realisations}
\end{figure}

Prior specification for the parameters governing the Multivariate Bernoulli detector (see Section~\ref{sec:prior}) is derived  from the following considerations. 
In the uncensored ($\delta=1$) single-risk ($m=1$) case,
the distribution of the time to event in the discrete case is a Time-Varying Geometric distribution \citep{Landau2019}
with time-varying success probability $\phi_t = \lambda(t\mid\theta)$,
i.e.\
\begin{equation} \label{eq:tvgeom}
    P(T  = t\mid \{\phi_l\}_l) =
    \begin{cases}
        \phi_t\, \prod_{l = 1}^{t - 1} (1 - \phi_l),
        \quad &t = 1,\dots,t_{\max} \\
        \prod_{l = 1}^{t - 1} (1 - \phi_l),
        \quad &t = t_{\max} + 1
    \end{cases}
\end{equation}
The Time-Varying Geometric is fully flexible in that it can represent any distribution
on $\{1,\dots,t_{\max}+1\}$
by appropriately choosing $\phi_t$
\citep*{Mandelbaum2007}.
It is analogous to the Piecewise Exponential distribution in continuous survival analysis \citep{Gamerman1991}, if we assume a change point model for the $\phi_t$. See Figure~\ref{fig:tvgeom_realisations} for widely varying realisations of the distribution for certain $\{\phi_t\}_t$.
This flexibility should be taken into account when inferring $\phi_t$. It relates to the potential lack of stability of unconstrained estimation of baseline hazards mentioned in Section~\ref{sec:intro}.
For instance, there is a separate parameter $\phi_t$ for each time point, but we might not have observed an event at each time point.
To avoid such overpameterization,
some subsequent $\phi_t$ can be assumed to be equal to each other, as in Figure~\ref{fig:tvgeom_realisations},
resulting in a change point model:
a change point is a time~$t$ for which $\phi_t\ne\phi_{t-1}$.
Moreover, given the flexibility of the Time-Varying Geometric, different combinations of $\{\phi_t\}$ can  lead to a satisfactory fit of a data set, leading to identifiability problems. As such, we impose prior regularization by 
a priori shrinking  the value of $\phi_t$
towards zero.
We further motivate our prior choice in the following simulation study.

We simulate $n=500$ times $t_i$ from \eqref{eq:tvgeom}
with $t_{\max} = 1000$ using two different settings for $\phi_t$.
We consider a scenario without change points with $\phi_t =  0.5$,
and a scenario with a single change point given by $\phi_t = 0.5$ for $t\leq 4$ and $\phi_t = 0.25$ for $t\geq 5$.
For this last scenario, we also consider the data after removal of observations with $t_i=7$, which we discuss in Section~\ref{sec:loc_cp} in relation to the prior constraints on change point location.
To understand how the prior on $\phi_t$ can affect inference on change points,
we compare two priors within a Bayesian change point model defined as follows:
we specify a uniform prior over all possible change point configurations
among the $\phi_t$.
Let $\phi_\ell^\star$ denote the unique value of $\phi_t$ over each time interval delimited by the change points.
Conditionally on a change point configuration,
we choose a prior on $\phi_\ell^\star$.
Then, the likelihood in \eqref{eq:tvgeom} completes the model.
We fit this model with $t_{\max} = \max_i t_i$,
such that $t_{\max}=9$ for the data with no change point and $t_{\max}=14$ for the data with a change point.
We compare posterior inference obtained with a uniform prior,
$\phi_\ell^\star\sim\mathcal{U}(0,1)$, 
and with a prior that shrinks the parameters towards zero, $\phi_\ell^\star\sim\mathrm{Beta}(0.01,1)$.

\begin{figure}
\centering
\includegraphics{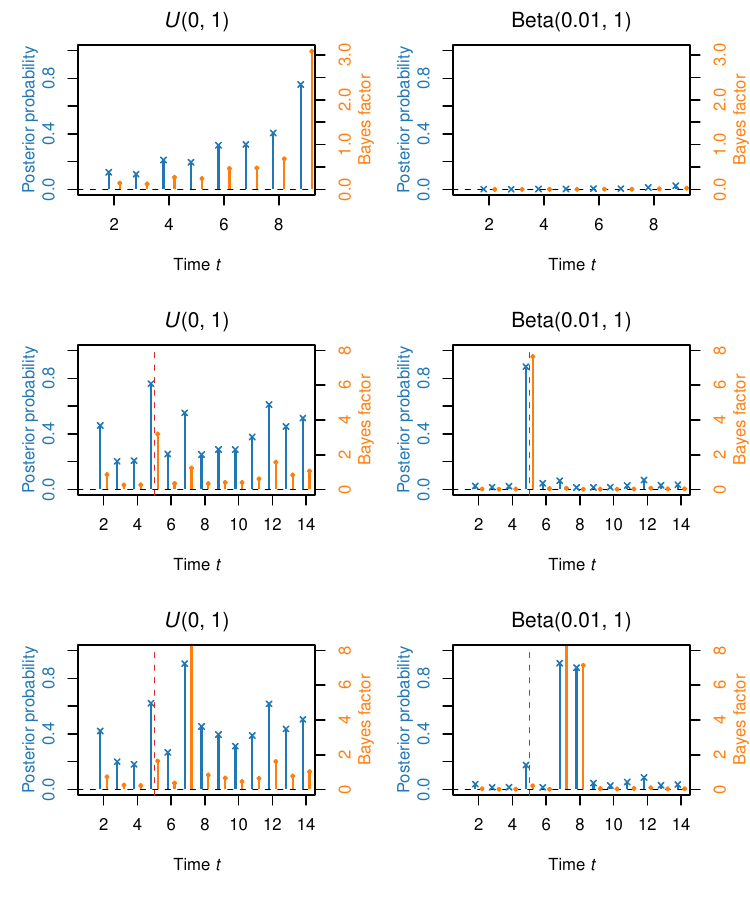}
\caption{Time-Varying Geometric simulation: Posterior probabilities ($\times$) and Bayes factors ($\bullet$) for the presence of a change point with a uniform prior (left column, $\phi_\ell^\star\sim\mathcal{U}(0,1)$)
and regularization towards zero (right column, $\phi_\ell^\star\sim\mathrm{Beta}(0.01,1)$) when simulating data without (top row) and with (middle and bottom rows) a change point.
The bottom row uses the data from the middle row without the observations with $t_i=7$.
The orange lines correspond to Bayes factors, some of which are outside the plotting range.
The dashed red lines are drawn in correspondence of the true change point.}
\label{fig:tvgeom}
\end{figure}

Figure~\ref{fig:tvgeom} shows the inference on change points.
The uniform prior  (left column) leads to the detection of too many change points, especially at larger $t$.
Regularization towards zero using $\phi_\ell^\star\sim\mathrm{Beta}(0.01,1)$
(right column)
yields more accurate posterior inference without spurious change points.

\subsection{Posterior computation using local-global MCMC}
\label{sec:mcmc}

To devise an MCMC scheme to perform posterior inference, we exploit the representation of the likelihood in \eqref{eq:likelihood} as a multinomial logistic regression \citep{tutz2016modeling}, using a data augmentation trick. This results in the availability of conjugate updates, leading to more efficient mixing and preventing, at the same time, large changes in the configuration of change points, resulting in more effective local moves.
The latent variables associated with the  data augmentation are highly correlated with the change points, and, as such, it is difficult to explore the change points space conditionally on the latent variables. To counterbalance this drawback, we also devise global moves of change points conditionally on the observed data. Such moves are based on ideas from the Bayesian nonparametric literature \citep{Dahl2005,Martinez2014,Creswell2020}.
Finally, from the MCMC output, we can derive Bayes factors to test for the presence of change points, using the Savage-Dickey ratio \citep{Dickey1971,Verdinelli1995} (see Web Appendix~\ref{ap:BF} for details).

We refer to the resulting hybrid algorithm as ``local-global MCMC'' borrowing the terminology from \citet{Samsonov2022}.
Here, we provide a brief explanation of our MCMC strategy in relation to previous work.
Web Appendix~\ref{ap:mcmc} details the algorithm.

\subsubsection{Local MCMC with data augmentation}

We exploit the data augmentation representation of a multinomial logistic regression in terms of Gumbel latent variables by \citet{McFadden1974} and \citet{Fruhwirth-Schnatter2007}.
Then, the augmented likelihood is Gaussian which enables convenient MCMC updates.
Importantly,
conditionally on the $z_{rt}$,
the $\alpha_{r\ell}^\star$
have a Gaussian prior such that they can be integrated out from the augmented posterior,
enabling efficient updates of $z_{rt}$ and $\gamma_t$ without having to specify Metropolis-Hastings proposals for $\alpha_{r\ell}^\star$.

More recently, other augmentations have been proposed in the literature \citep*[see, for instance,][]{Held2006,Fruhwirth-Schnatter2010,Polson2013,Linderman2015}.
We do not opt for them because they do not provide a convenient augmented likelihood in the presence of multiple risks.

\subsubsection{Global MCMC}

Augmented data can strongly reflect the change points of the current state of the MCMC chain, resulting in local MCMC updates to the change point parameters $z_{rt}$ and $\gamma_t$. Therefore, we also consider MCMC moves without data augmentation, i.e.\ based on the original data, to enable more global change point updates and explore better posterior space.

Specifically,
we exploit the fact that change points induce a partition of time into intervals and apply ideas from Bayesian nonparametrics \citep{Dahl2005,Martinez2014,Creswell2020} to deal with non-conjugate updates.
This allows for more global moves at the cost of having to specify
Metropolis-Hastings proposals for $\alpha_{r\ell}^\star$.
Alternating between local and global MCMC steps allows for better mixing and convergence of the algorithm.

We demonstrate the performance of our approach in simulation studies.
Web Appendix~\ref{ap:simul} presents simulation studies with a wide range of scenarios and comparison with alternative models. We find that the Multivariate Bernoulli detector generally results in the most accurate estimation. Prior shrinkage of baseline hazards can result in estimation bias, which is a common feature of Bayesian shrinkage priors and the bias-variance trade-off they induce \citep[e.g.][]{Polson2019}.

\section{Application to \texorpdfstring{ICU length of stay}{ICU length of stay}}
\label{sec:application}

\subsection{Data description and analysis}
\label{sec:data_description}

We apply our model to
data on
ICU stays
from the
MIMIC\nobreakdash-IV database \citep{Johnson2023}
with length of stay as outcome \citep{Meir2023}.

See Web Appendix~\ref{ap:mimic} for a detailed data description.
We consider $m=3$ competing risks: discharge to home, transfer to another medical facility and in-hospital mortality.
Length of stay is recorded in days with the longest uncensored time being 28 days.
We analyse $n=25159$ ICU stays with $17357$ discharged to home, $5379$ transferred, $1529$ deaths and $894$ censored.
We include the following covariates: demographics, variables related to the ICU stay (e.g.\ whether it is a repeat admission) and lab tests from the first day. Most covariates are categorical with two or more levels. Representing them as dummy variables leads to a total of  $p=36$ predictors.

We fit our model with
$t_{\max} = \max_i t_i = 28$ days
using 200000 MCMC iterations, discarding the first 50000 as burn-in.
The trace plots in Web Figure~\ref{fig:mimic_mcmc} suggest satisfactory convergence.

%%% Manual adjustments due to figure placement:
\pagebreak
\vspace*{.1in}
\subsection{Posterior inference on the baseline hazards}

\begin{figure}
\centering
\includegraphics[width=\textwidth]{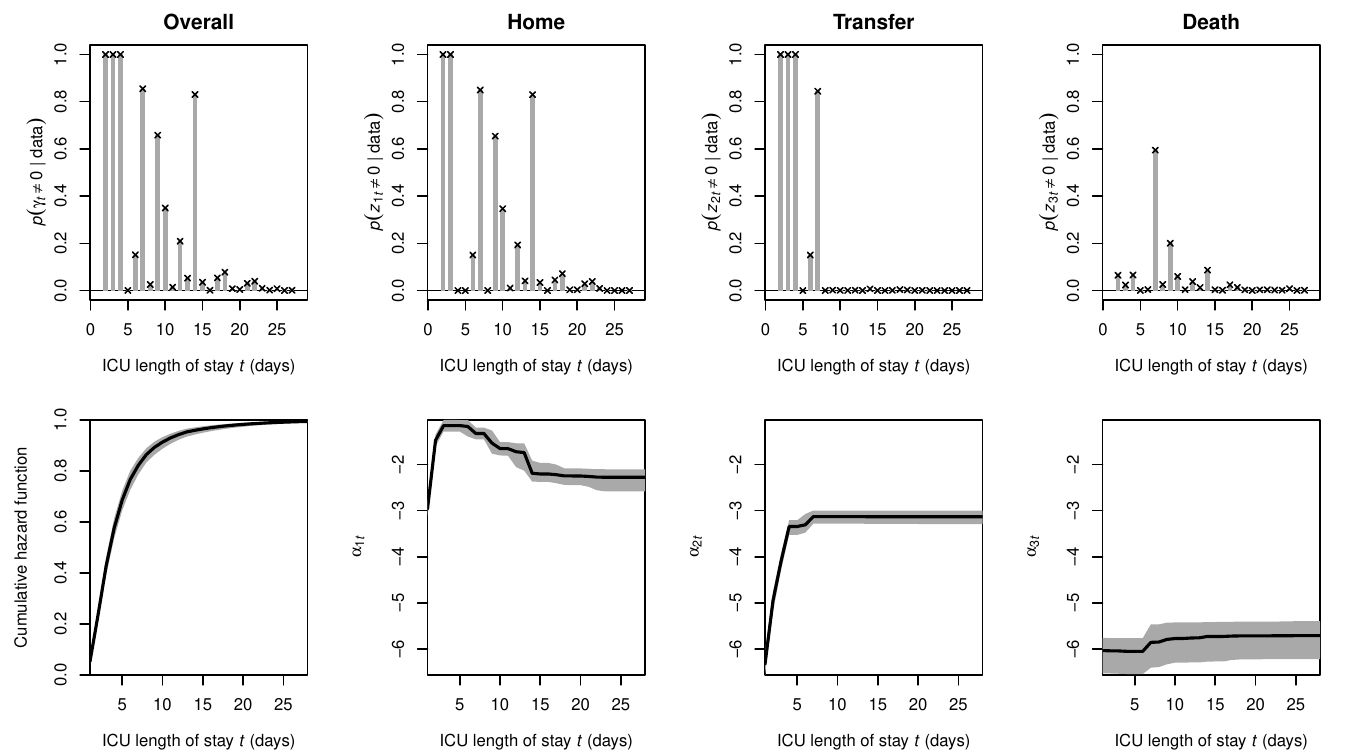}
\caption{ICU data: Posterior inference on the overall (left column) and cause-specific (other columns) baseline hazards. The top row displays the posterior probabilities  for the presence of a change point. The bottom row shows posterior inference for the cumulative hazard function for $\bm x_i = \bm 0$, and the baseline hazard parameter $\alpha_{rt}$.
Black lines represent posterior means and shaded areas correspond to 95\% credible intervals.}
\label{fig:mimic_hazard}
\end{figure}

The posterior probability of absence of overall change points is zero as well as the Bayes factor (see  Web Appendix~\ref{ap:BF}).
%which suggests the presence of change points.
Figure~\ref{fig:mimic_hazard} summarizes inference on the baseline hazards (see Web Figure~\ref{fig:mimic_hazard_BF} for corresponding Bayes factors).
The hazard functions
differ markedly between risks: the hazard of discharge to home is high in the first two weeks, but not on the first day in the ICU. The hazard of a transfer to another medical facility is lowest during the first few days. Finally, the hazard function for in-hospital mortality does not vary substantially across the length of stay.

\begin{figure}
\centering
\includegraphics[width=0.5\textwidth]{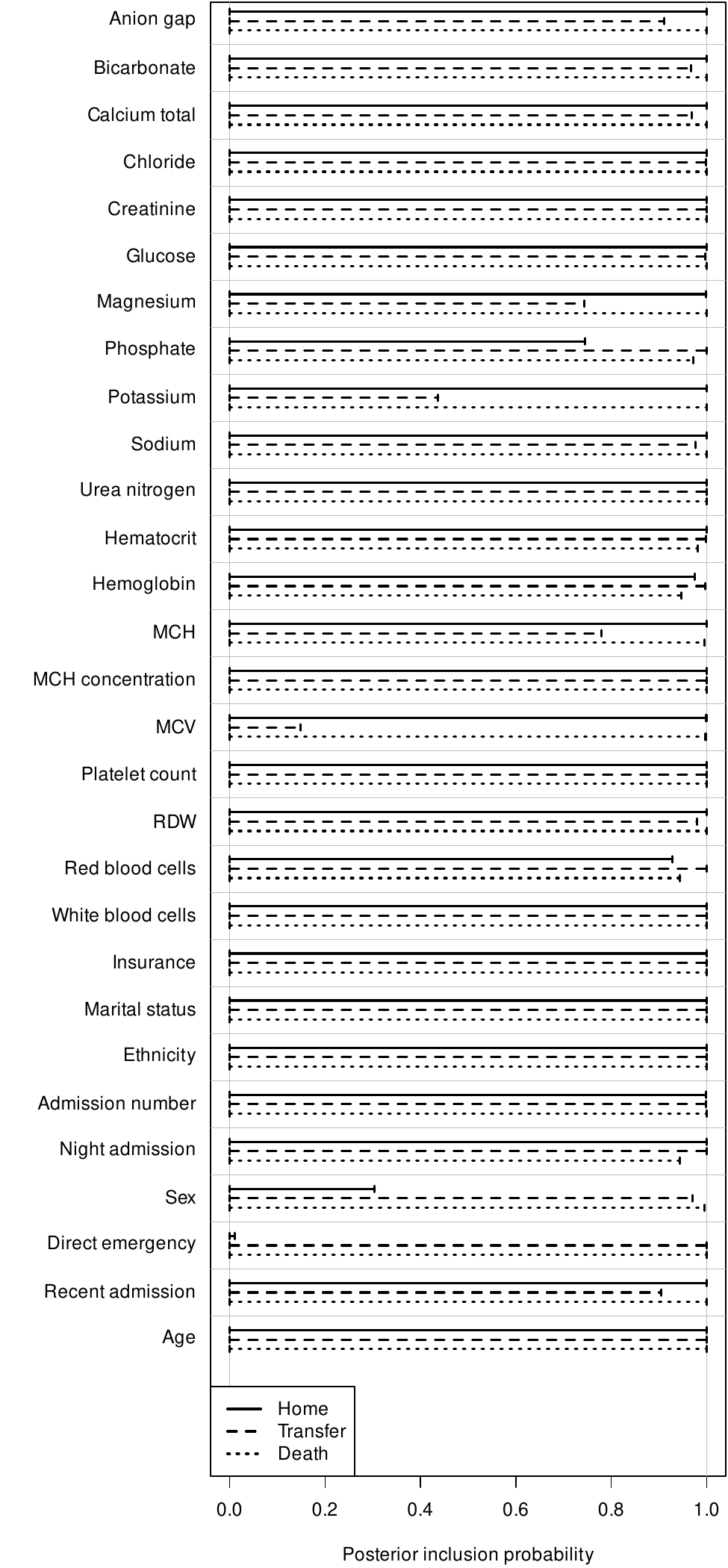}
\caption{ICU data: Posterior inclusion probabilities for each covariate and risk. MCH stands for mean cell hemoglobin, MCV for mean corpuscular volume and RDW for red blood cell distribution width.
}
\label{fig:mimic_inc_prob}
\end{figure}

\begin{figure}
\centering
\includegraphics[width=0.5\textwidth]{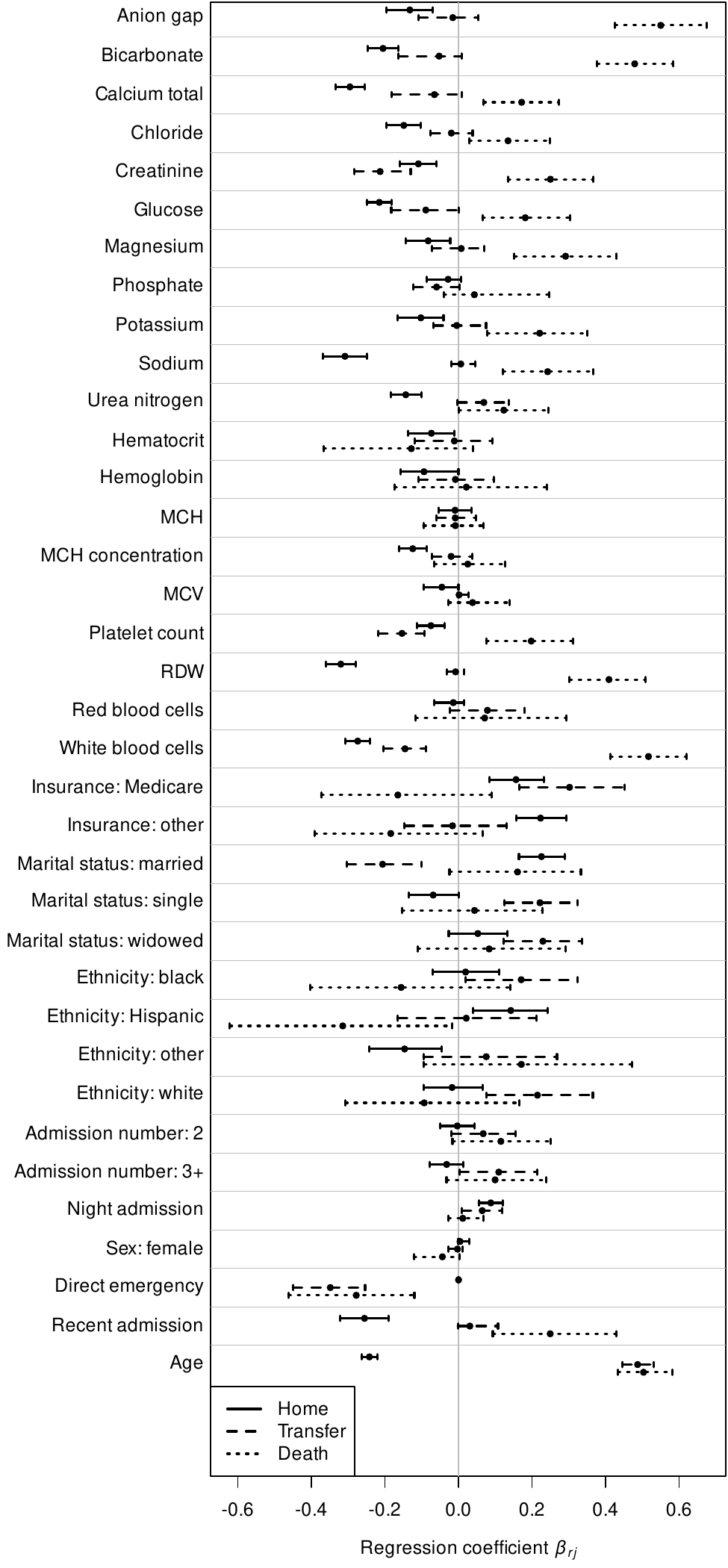}
\caption{ICU data: Posterior means (dot) and 95\% credible intervals (lines) of the regression coefficients for each risk.
The categorical predictors are coded as dummy variables as detailed in Web Appendix~\ref{ap:mimic}. MCH stands for mean cell hemoglobin, MCV for mean corpuscular volume and RDW for red blood cell distribution width.
}
\label{fig:mimic_regression}
\end{figure}

\FloatBarrier

\subsection{Posterior inference on the regression coefficients}

The regression coefficients $\beta_{rj}$ capture cause-specific covariate effects on length of stay.
The spike-and-slab prior provides explicit inference on whether there is a covariate effect through the posterior probability of $\beta_{rj}\ne 0$.
Posterior inclusion probabilities for each risk are shown in Figure~\ref{fig:mimic_inc_prob}.
In 
Figure~\ref{fig:mimic_regression}, we report posterior inference on regression coefficients.
Finally, we remark that the results on the baseline hazards and covariate effects are in line with those obtained in \citet{Meir2023}.

\section{Comparison with other models}
\label{sec:compare}

We compare our results on the ICU data to those obtained from maximum likelihood estimation and a more recent alternative, namely the model by \citet{King2021}.

\subsection{Maximum likelihood estimation}
\label{sec:mle}

We maximize the likelihood in \eqref{eq:likelihood}  using the R package \texttt{nnet} \citep{Venables2002} as the likelihood is equivalent to a multinomial logistic regression \citep{tutz2016modeling}.
The resulting inference is shown in Web Figures~\ref{fig:mimic_nnet_hazard} and \ref{fig:mimic_nnet_regression}.
Estimates of $\bm\alpha_t$ are less smooth than for our model, but are otherwise similar.
Estimates of $\bm\beta_r$ are in agreement with ours.

\subsection{Semi-parametric model by \texorpdfstring{\citet{King2021}}{King and Weiss (2021)}}
\label{sec:King2021}

We also compare our model with the Bayesian semi-parametric model by \citet{King2021}, which also involves multiple risk and a flexible model for the hazard function.
For simplicity of explanation in what follows, we denote with $x_{ij}$ a continuous covariate $j$ for individual $i$ and with $d_{ik}$ a dummy variable corresponding to a level of a categorical covariate. 
\citet{King2021}
specify a multinomial logit model for discrete survival analysis with competing risks
with
$\eta_{irt} = \alpha_{rt} + \sum_{j=1}^{p^c} f_{\beta r j}(x_{ij}) + \sum_{k=1}^{p^d} \beta_{rk}\, d_{ik}$
where $p^c$ and $p^d$ denote the number of continuous and dummy variables, respectively.
Moreover, 
$\alpha_{tr} = \beta_{0r} + f_{\alpha r}(t)$
for intercepts $\beta_{0r}$, and functions $f_{\alpha r}$ and $f_{\beta rj}$
which are object of inference. Note that in their approach, \citet{King2021} include every level of a categorical covariate.
The functions $f_{\alpha r}$ and $f_{\beta rj}$
are inferred using a Gaussian Markov random field prior.
For prior specification and parameter choice, we follow the recommendations in Appendix~C of \citet{King2021} for uninformative priors.
We fit the model using the R package
\texttt{brea} \citep{King2017}
using 50000 burn-in MCMC iterations
followed by
200000 recorded iterations.

The resulting inference is shown in Web Figures~\ref{fig:mimic_brea_hazard}, \ref{fig:mimic_brea_regression_age} and \ref{fig:mimic_brea_regression}.
The estimates of baseline hazards are in line with our model, though not as smooth.
The non-linear covariate
effect of age is consistent with the linear effect from our model, but positive association of age and transfer hazard only appears at an older age. The other covariate effects are also
similar to the results from our model. See  Figure~\ref{fig:mimic_regression}.

\section{Discussion}
\label{sec:discussion}

In this work, we focus on the estimation of the hazard function of competing risks  in the context  of discrete survival. We assume a change point model for the hazard function, with cause-specific change points, introducing dependence among change point locations across risks. In our approach, both number and location of change points are random. We refer to our model as   
Multivariate Bernoulli detector. Dependence across risks provides an attractive way for regularization of baseline hazards since changes to an individual's condition across time might affect multiple cause-specific hazards simultaneously. Our approach is widely applicable and interpretable. The data augmentation enables the use of any prior on regression coefficients making the MCMC updates more efficient.
The simulation study and the real data application show that posterior inference on change points with dependence across risks is effective, with favorable comparisons with a frequentist approach and the Bayesian semi-parametric model by \citet{King2021}.
 
The proposed model can be easily extended to accommodate more complex scenarios, for example, inclusion of 
recurrent event processes as outcome \citep[see, e.g.,][]{King2021}, of time-varying covariates or semi-competing risk structure.
In this work, we employ the multinomial logit model which is a popular choice for the analysis of discrete competing risks. It closely relates to multinomial logistic regression and offers computational advantages. Nevertheless, the Multivariate Bernoulli detector can be used with other likelihoods, such as multinomial probit models or multiple time series. Finally, we note that we could apply the same computational strategy even for change points models in continuous time by restricting the split points to the
locations of the events.

%  The \backmatter command formats the subsequent headings so that they
%  are in the journal style.  Please keep this command in your document
%  in this position, right after the final section of the main part of 
%  the paper and right before the Acknowledgements, Supporting Information (Supplementary %  Materials),   and References sections. 

\backmatter

%  This section is optional.  Here is where you will want to cite
%  grants, people who helped with the paper, etc.  But keep it short!

\section*{Acknowledgements}

This work is supported by the Singapore Ministry of Health's National Medical Research Council under its Open Fund - Young Individual Research Grant (OFYIRG19nov-0010). A.G. has been partially supported by MUR - Prin 2022 - Grant no. 2022CLTYP4, funded by the European Union - Next Generation EU.

\section*{Supplementary Materials}

Web Appendices and Figures referenced in Sections~\ref{sec:model}, \ref{sec:application} and \ref{sec:compare} are available with in this paper's Supplementary Materials.
The code to implement the model is available at \url{https://github.com/willemvandenboom/mvb-detector}.

\section*{Data availability statement}

The data that support the findings of this study are available from PhysioNet. Restrictions apply to the availability of these data, which were used under license for this study. Data are available at \url{physionet.org} with the permission of PhysioNet.

%  Here, we create the bibliographic entries manually, following the
%  journal style.  If you use this method or use natbib, PLEASE PAY
%  CAREFUL ATTENTION TO THE BIBLIOGRAPHIC STYLE IN A RECENT ISSUE OF
%  THE JOURNAL AND FOLLOW IT!  Failure to follow stylistic conventions
%  just lengthens the time spend copyediting your paper and hence its
%  position in the publication queue should it be accepted.

%  We greatly prefer that you incorporate the references for your
%  article into the body of the article as we have done here 
%  (you can use natbib or not as you choose) than use BiBTeX,
%  so that your article is self-contained in one file.
%  If you do use BiBTeX, please use the .bst file that comes with 
%  the distribution.  In this case, replace the thebibliography
%  environment below by 
%
%  \bibliographystyle{biom} 
% \bibliography{mybibilo.bib}

\bibliographystyle{biom}
\bibliography{Ref_discretetimesurvival}

\label{lastpage}

\end{document}

\title{
Supplementary Materials for\\ ``The Multivariate Bernoulli detector:\\ Change point detection in discrete survival analysis''\\
by Willem van den Boom, Maria~De~Iorio,\\ Fang Qian and Alessandra Guglielmi
}

\date{}

\maketitle
\appendix

\section{ICU data description}
\label{ap:mimic}

The intensive care unit (ICU) data analyzed in Section~\ref{sec:application} of the main manuscript
are taken from \citet{Meir2023}.
The data are
from the
Medical Information Mart for Intensive Care IV (MIMIC\nobreakdash-IV) database version~2.0 \citep{Johnson2022,Johnson2023} which is available from PhysioNet \citep{Goldberger2000} at \url{https://physionet.org/content/mimiciv/2.0}.
MIMIC\nobreakdash-IV contains critical care data from patients admitted to ICUs at the Beth Israel Deaconess Medical Center,
which is a tertiary academic medical centre in Boston, Massachusetts.

The analysis considers ICU stays from 2014 to 2020 with admission type `direct emergency', i.e.\ hospital admission directly into the ICU, or `emergency ward', i.e.\ admission to the ICU from the emergency ward.
Furthermore, if a patient has multiple ICU stays, then only the last stay is considered.
Ultimately,
the data consist of
$n=25159$~ICU stays.
% \footnote{
% The number of ICU stays is 11 less than the number mentioned in \citet{Meir2023}. Based on the code \citep{Meir2023b} provided by \citeauthor{Meir2023}, the 11 ICU stays are excluded because their discharge time precedes their admission time in MIMIC-IV, which is likely an administrative error: see
% \url{https://github.com/MIT-LCP/mimic-code/issues/209}.
% }
The main causes of termination of ICU stays are
discharge to home
($17357$~stays / $69.0\%$),
transfer to another medical facility
($5379$~stays / 21.4\%)
and in-hospital mortality
($1529$~stays / 6.1\%).
Furthermore, $1.0\%$ of patients leave the ICU against medical advice, which are considered as censored.
Additionally, \citet{Meir2023} censor length of stay at 28 days.
In total, 894 stays ($3.6\%$) are censored.

Covariates derive from
demographics, variables related to the ICU stay and lab tests.
The demographics are (i) standardised age; (ii) sex; (iii) ethnicity; (iv) marital status; (v)~insurance type.
The variables related to the ICU stay
are (i) admission type;
(ii) whether the admission occurred at night, i.e.\ between 8am and 8pm;
(iii) whether the patient was also admitted during the previous 30 days (\emph{recent admission});
(iv) the number of emergency admissions for the patient (\emph{admission number}).
For instance, the admission number is equal to one if the ICU stay is not a repeat admission and equal to two if it is the patient's second emergency admission.
We also include
20 lab measurement
that are the
most commonly
recorded within the first 24 hours after admission.
Lab tests are coded as binary variables with 1 indicating an abnormal result and 0 otherwise.
Categorical covariates are listed in Web Table~\ref{tab:baseline}. Note that \emph{age} is the only continuous covariate. 
Categorical covariates are included in the regression term through the standard dummy variable representation, fixing a category as baseline (see Web Table~\ref{tab:baseline}). 
This results in $p=36$ predictors.
Summary statistics of the data are presented in Tables~4 and 5 of \citet{Meir2023}.

\begin{table}
    \centering

    \caption{Categorical covariates.}
    \label{tab:baseline}
    \vspace{0.3cm}
    \begin{threeparttable}
    \begin{tabular}{l|lllll}
    \textbf{Covariate} & \textbf{Baseline} & \multicolumn{4}{l}{\textbf{Other categories}} \\
    \hline \\
    Demographics \\
    \emph{Sex}     & Male & Female \\
    \emph{Ethnicity}     & Asian & White & Black & Hispanic & Other \\
    \emph{Marital status} & Divorced & Single & Married & Widowed \\
    \emph{Insurance type} & Medicaid & Medicare & Other \\
    \hline \\
    Related to ICU stay \\
    \emph{Admission type} & Emergency ward & \multicolumn{4}{l}{Direct emergency} \\
    \emph{Night admission} & No & Yes \\
    \emph{Recent admission} & No & Yes \\
    \emph{Admission number} & 1 & 2 & 3 or more \\ \hline \\
    Lab tests \\
    \emph{Anion gap} & Normal & \multicolumn{4}{l}{Abnormal} \\
    \emph{Bicarbonate} & Normal & \multicolumn{4}{l}{Abnormal} \\
    \emph{Calcium total} & Normal & \multicolumn{4}{l}{Abnormal} \\
    \emph{Chloride} & Normal & \multicolumn{4}{l}{Abnormal} \\
    \emph{Creatinine} & Normal & \multicolumn{4}{l}{Abnormal} \\
    \emph{Glucose} & Normal & \multicolumn{4}{l}{Abnormal} \\
    \emph{Magnesium} & Normal & \multicolumn{4}{l}{Abnormal} \\
    \emph{Phosphate} & Normal & \multicolumn{4}{l}{Abnormal} \\
    \emph{Potassium} & Normal & \multicolumn{4}{l}{Abnormal} \\
    \emph{Sodium} & Normal & \multicolumn{4}{l}{Abnormal} \\
    \emph{Urea nitrogen} & Normal & \multicolumn{4}{l}{Abnormal} \\
    \emph{Hematocrit} & Normal & \multicolumn{4}{l}{Abnormal} \\
    \emph{Hemoglobin} & Normal & \multicolumn{4}{l}{Abnormal} \\
    \emph{MCH} & Normal & \multicolumn{4}{l}{Abnormal} \\
    \emph{MCH concentration} & Normal & \multicolumn{4}{l}{Abnormal} \\
    \emph{MCV} & Normal & \multicolumn{4}{l}{Abnormal} \\
    \emph{Platelet count} & Normal & \multicolumn{4}{l}{Abnormal} \\
    \emph{RDW} & Normal & \multicolumn{4}{l}{Abnormal} \\
    \emph{Red blood cells} & Normal & \multicolumn{4}{l}{Abnormal} \\
    \emph{White blood cells} & Normal & \multicolumn{4}{l}{Abnormal} \\
    \hline
    \end{tabular}
    \vspace{0.3cm}
    \begin{tablenotes}
        \item MCH stands for mean cell hemoglobin, MCV for mean corpuscular volume and RDW for red blood cell distribution width.
    \end{tablenotes}
    \end{threeparttable}

\end{table}

\section{Groupwise variable selection prior}
\label{ap:var_sel}

As discussed above, most of the available covariates are categorical, some of which characterised by more than two
levels, e.g.\ \emph{ethnicity}.
When performing variable selection, we aim to either keep all the dummy variables corresponding to the same categorical covariate in the model or to exclude all of them.
Specifically,
let
$S_l$ denote the set of dummy variables corresponding to categorical covariate $l$.
We propose the following modification of the prior on the corresponding regression coeficients from Section~\ref{sec:prior} for groupwise variable selection. 

Let $p_l=|S_l|$ and let $\bm\beta_{rS_l}$ denote the $p_l$-dimensional subvector of $\bm\beta_r$ corresponding to variables in $S_l$.
For each risk $r$ and group $l$ independently, we assume
\[\bm\beta_{rS_l}\sim \pi_\beta\, \mathcal{N}_{p_l}(\bm 0,\, \sigma^2_\beta\, I_{p_l}) + (1-\pi_\beta)\, \delta_{\bm 0}\]
where
$\mathcal{N}_{p_l}(\bm 0, \sigma^2_\beta\, I_{p_l})$
is a $p_l$-dimensional Gaussian distribution with mean zero and covariance matrix $\sigma^2_\beta\, I_{p_l}$,
$\bm 0$ is the
zero vector
and $\delta_{\bm 0}$ denotes a point mass at $\bm 0$.
Thus,
a priori,
the elements of $\bm\beta_{rS_l}$ are all non-zero (resp.\ zero) simultaneously with probability
$\pi_\beta$ (resp.\ $1-\pi_\beta$).
The hyperprior on $\pi_\pi$ remains $\pi_\beta \sim \mathcal{U}(0, 1)$ as in Section~\ref{sec:prior} of the main manuscript.

\section{Bayes factor for change points}
\label{ap:BF}

We now describe a strategy to test for the presence of change points. In particular, we provide a strategy to compute the Bayes factor $B$ for $K=0$ (model~$\mathcal{M}^\star$, denoting no change points) versus a model with $K\sim p(K)$ (model~$\mathcal{M}$) \citep*[see][for a similar example of such model comparison]{Legramanti2022}.
$B$ is readily computed from Markov chain Monte Carlo (MCMC)
output. Let $\bm y$ denote all the observed data. Then
\[
    B = \frac{p(\bm y\mid \mathcal{M}^\star)}{p(\bm y\mid \mathcal{M})}
    = \frac{p(\bm y\mid K = 0)}{p(\bm y)}
    = \frac{p(\bm y, K = 0)}{p(\bm y)\, p(K=0)}
    = \frac{p(K = 0\mid \bm y)}{p(K=0)}
\]
where the last ratio is the Savage-Dickey ratio \citep{Dickey1971,Verdinelli1995}.
Here,
${p(K = 0\mid \bm y)}$ is readily estimated by the MCMC sample frequency of $K=0$
while $p(K=0)$ is available from the prior.

The scheme can be employed as long as $p(K=0)$ is not too small as it would lead to unstable estimation of $B$ when also $p(K = 0\mid \bm y)$ is small (i.e.\ the MCMC chain visits $K=0$ only rarely).
This is not the case in our scenario as $K \sim\mathrm{Geo}_{|\mathcal{T}|}(\pi_K)$
such that $p(K=0) > \pi_K = 0.5$.

\section{Local-global MCMC algorithm}
\label{ap:mcmc}

Here, we
provide
further details on the MCMC, including the data augmentation from \citet{Fruhwirth-Schnatter2007}, and the methodology from Bayesian nonparametrics based on \citet{Dahl2005}, \citet{Martinez2014} and \citet{Creswell2020}.

Let $\bm\theta = (\theta_1,\dots,\theta_n)$.
Let $\bm y$ denote all the observed data.
Then, the likelihood for all observations is
\begin{equation} \label{eq:likelihood2}
p(\bm y\mid \bm\theta) = \prod_{i=1}^n \left[
\lambda_{r_i}(t_i\mid\theta_i)^{\delta_i}\,
    \prod_{l=1}^{t_i - \delta_i} \{1 - \lambda(l\mid\theta_i)\} \right]
\end{equation}

\subsection{Local MCMC with data augmentation}

For the data augmentation,
we consider
the
following
Gaussian mixture approximation of the standard Gumbel density:
$\exp(-u - e^{-u}) \approx \sum_{c = 1}^{10} w_c\, \mathcal{N}(u\mid \xi_c,\, s_c^2)$
where the weights $w_c$, means $\xi_c$ and variances $s_c^2$ of each component $c$ are given in Table~1 of \citet{Fruhwirth-Schnatter2007}.
Conditionally on the linear predictors $\eta_{irt}$,
we sample the augmented data, consisting of a latent variable $u_{irt}$ and a component indicator $c_{irt}$,
according to Step~(b) in Section~3.2 of \citet{Fruhwirth-Schnatter2007}
as detailed in Algorithm~\ref{alg:augmentation}.
Then, the augmented likelihood follows as
\[
    p(\bm y\mid \bm u,\bm c, \bm\theta) = \prod_{r=1}^m \prod_{i=1}^n \prod_{l=1}^{t_i} \mathcal{N}(u_{irl}\mid \eta_{irl} + \xi_{c_{irl}},\, s_{c_{irl}}^2)
\]
Its Gaussianity enables convenient MCMC updates for the parameters constituting $\bm\theta$.

\begin{algorithm}
\caption{Sampling of the augmented data $u_{irt}$ and $c_{irt}$. \label{alg:augmentation}}
For $i = 1,\dots,n$; for $l = 1,\dots,t_i$:
\begin{enumerate}
    \item
    Sample $U_{il}\sim\mathcal{U}(0,1)$.
    \item
    For $r=1,\dots,m$,
    sample $V_{irl}\sim\mathcal{U}(0,1)$
    and set
    \[
        u_{ir{l}} = -\log\left\{-\frac{\log(U_{i{l}})}{1 + \sum_{\rho=1}^m \exp(\eta_{i\rho{l}})} - \mathds{1}\left[ \begin{array}{rl}
             & {l}\ne t_i \\
        \text{or} & \delta_i = 0 \\
        \text{or} & r_i\ne r
        \end{array} \right]
        \frac{\log(V_{ir{l}})}{\exp(\eta_{ir{l}})}
        \right\}
    \]
    \item \label{step:sample_c}
    Sample $c_{ir{l}}\in\{1,\dots,10\}$ according to
    \[
        P(c_{ir{l}} = c) \propto \frac{w_c}{s_c}\exp\left\{
            -\frac{1}{2s_c^2}(u_{ir{l}} - \eta_{ir{l}} - \xi_c)^2
        \right\}
    \]
\end{enumerate}
\end{algorithm}

Firstly, we perform a split-merge-shuffle step based on \citet{Martinez2014} to update $\gamma_t$ with $\alpha_{r\ell}^\star$ integrated out from the augmented likelihood.
Here, \emph{split} refers to the addition of an overall change point that splits a time interval, i.e.\ constant baseline hazard, into two.
\emph{Merge} refers to the deletion of a change point that merges two adjacent time intervals.
Lastly, a \emph{shuffle} is the relocation of a change point such that the lengths of its adjacent time intervals change.

For the description of the MCMC step, we introduce additional notation.
The overall change points marked by $\gamma_t$ partition the time points $\{1, \ldots, t_{\max}\}$ into $K+1$ time intervals. Denote the number of time points in the ${\ell}$-th time interval by $n_{\ell}$.
Due to conjugacy, we can integrate out $\alpha_{rt}$ from the augmented likelihood to obtain a closed form expression for $p(\bm y\mid \bm u,\bm c, \bm\theta\setminus\bm\alpha) = {p(\bm y\mid \bm u,\bm c, \bm\zeta)}$
where
$\bm\zeta = \bm\theta\setminus\bm\alpha$ only consists of change point indicators and regression coefficients.
We detail the split-merge-shuffle step in Algorithm~\ref{alg:update_gamma_augmented}.

\begin{algorithm}
\caption{Split-merge-shuffle step for $\gamma_t$ based on the augmented likelihood. \label{alg:update_gamma_augmented}}
\begin{enumerate}
    \item
    Perform a split or a merge step:
    \begin{enumerate}
        \item
        With probability $\mathds{1}[K=0] + \frac{1}{2}\mathds{1}[0<K<t_{\max} - 1]$,
        perform a split step:
        \begin{enumerate}
            \item
            Sample a time interval ${\ell}'$ uniformly from $\{{\ell}\mid n_{\ell} \geq 2\}$.
            \item
            Split time interval ${\ell}'$ by setting $\gamma_{t'} = 1$ for a $t'$ sampled uniformly from $\{2 + \sum_{\kappa = 1}^{{\ell}'-1} n_\kappa,\,3 + \sum_{\kappa = 1}^{{\ell}'-1} n_\kappa,\,\dots,\, \sum_{\kappa = 1}^{{\ell}'} n_\kappa\}$.
            \item
            Sample corresponding cause-specific change points $\bm z_{t'}'$
            from the prior $p(\bm z_{t'}\mid \gamma_{t'})$.
            Let $\bm\zeta'$ denote the resulting parameter where the other change points and $\bm\beta_{r}$ are left unchanged.
            \item
            Set $\bm\zeta=\bm\zeta'$ with probability
            \[
                \min\left\{1,\, \frac{2^{\mathds{1}[K > 0]}\, |\{{\ell}\mid n_{\ell} \geq 2\}|\, (n_{{\ell}'} - 1)\,p(\bm\zeta')\, p(\bm y\mid \bm u,\bm c, \bm\zeta')}{2^{\mathds{1}[K < t_{\max} - 2]}\, (K + 1)\,p(\bm z'_{t'}\mid \gamma_{t'})\,p(\bm\zeta)\, p(\bm y\mid \bm u,\bm c, \bm\zeta)}\right\}
            \]
        \end{enumerate}
        \item
        Otherwise, perform a merge step:
        \begin{enumerate}
            \item
            Sample a time interval ${\ell}'$ uniformly from $\{1,\dots,K\}$.
            \item
            Let $\bm\zeta'$ denote the parameter resulting from merging time intervals ${\ell}'$ and $({\ell}'+1)$, and the corresponding removal of change points.
            \item
            Let $t_\text{old} = 1 + \sum_{\kappa=1}^{{\ell}'} n_\kappa$.
            Set $\bm\zeta=\bm\zeta'$ with probability
            \[
                \min\left\{1,\, \frac{
                2^{\mathds{1}[K < t_{\max} - 1]}\, K\,p(\bm z_{t_\textnormal{old}}\mid \gamma_{t_\textnormal{old}})\,p(\bm\zeta')\, p(\bm y\mid \bm u,\bm c, \bm\zeta')
                }{
                2^{\mathds{1}[K > 1]}\, |\{{\ell}\mid n_{\ell}' \geq 2\}|\, (n_{{\ell}'}' - 1)\,p(\bm\zeta)\, p(\bm y\mid \bm u,\bm c, \bm\zeta)}\right\}
            \]
            where $n_{{\ell}'}'$ is the duration of the merged time interval.
        \end{enumerate}
    \end{enumerate}
    \item
    If $K > 0$, perform a shuffle step:
    \begin{enumerate}
        \item
        Sample a time interval ${\ell}'$ uniformly from $\{1,\dots,K\}$. Let $t_\text{old}=1 + \sum_{\kappa=1}^{{\ell}'} n_\kappa$.
        \item
        Sample a new overall change point $t'$ uniformly from\\  % Manual line break ---------------------------------------
        $\{{t_\text{old} - n_{{\ell}'}},\,t_\text{old} - n_{{\ell}'} + 1,\,\dots,\, {t_\text{old} + n_{{\ell}'+1} - 1}\}$.
        \item
        Let $\bm\zeta'$ denote the parameter resulting from setting
        $\gamma'_{t'}=1$, $\bm z'_{t'} = \bm z_{t_\text{old}}$ and,
        if $t'\ne t_\text{old}$,
        $\gamma'_{t_\text{old}} = 0$
        and
        $\bm z_{t_\text{old}} = \bm 0$.
        \item
        Set $\bm\zeta=\bm\zeta'$ with probability
        \[
            \min\left\{1,\, \frac{
           p(\bm\zeta')\,p(\bm y\mid \bm u,\bm c, \bm\zeta')
            }{
            p(\bm\zeta)\,p(\bm y\mid \bm u,\bm c, \bm\zeta)}\right\}
        \]
    \end{enumerate}
\end{enumerate}
\end{algorithm}

Metropolis-Hastings steps for $\bm z_{t}$, $\alpha_{r\ell}^\star$ and $\bm\beta_r$ follow from the Gaussian augmented likelihood as detailed in Algorithms~\ref{alg:update_z_augmented}, \ref{alg:update_alpha} and \ref{alg:update_beta}, respectively.
We state the update for $\bm\beta_r$ in Algorithm~\ref{alg:update_beta} for the groupwise variable selection prior in Web Appendix~\ref{ap:var_sel}. The update for the non-grouped prior in Section~\ref{sec:prior} of the main manuscript can be obtained by assigning each predictor to its own group $S_l$, i.e.\ $S_l=\{l\}$ for $l=1,\dots,p$.

\begin{algorithm}
\caption{Metropolis-Hastings step for $\bm z_{t}$ based on the augmented likelihood. \label{alg:update_z_augmented}}
If $K>0$, update $\bm z_t$ corresponding to a randomly sampled time interval:
\begin{enumerate}
    \item
    Sample ${\ell}'$ uniformly from $\{2,\dots,K+1\}$. Let $t' = 1 + \sum_{\kappa=1}^{{\ell}' - 1} n_{\ell}$.
    \item
    Sample cause-specific change points $\bm z_{t'}'$
    from the prior $p(\bm z_{t'}\mid \gamma_{t'})$.
    Let $\bm\zeta'$ denote the resulting parameter where the other change points and $\bm\beta_{r}$ are left unchanged.
    \item
    Set $\bm\zeta=\bm\zeta'$ with probability
    \[
        \min\left\{1,\, \frac{
       p(\bm y\mid \bm u,\bm c, \bm\zeta')
        }{
        p(\bm y\mid \bm u,\bm c, \bm\zeta)}\right\}
    \]
\end{enumerate}
\end{algorithm}

\begin{algorithm}
\caption{Gibbs sampling step for $\alpha_{r\ell}^\star$ based on the augmented likelihood. \label{alg:update_alpha}}
For $r=1,\dots,m$, update $\alpha_{r\ell}^\star$:
\begin{enumerate}
    \item
    Let $S^r_\ell = \{t\mid \ell - 1 \leq \sum_{{l}=2}^t z_{r{l}} < \ell \}$
    denote the set of times in the $\ell$-th cause-specific time interval for competing risk $r$.
    \item
    For $\ell=1,\,\dots,\,1 + \sum_{{l}=2}^{t_{\max}} z_{r{l}}$, sample $\alpha^\star_{r\ell} = \alpha_{rt}$, $t\in S^r_\ell$, corresponding to the $\ell$-th time interval from its full conditional:
    \[
        \alpha^\star_{r\ell} \mid \bm u,\bm c, \bm\zeta \sim \mathcal{N}\left\{
            \sigma^2_{\alpha r\ell}\left( \frac{\mu_\alpha}{\sigma^2_\alpha} + \sum_{{l}\in S^r_\ell} \sum_{\{i\mid {l}\leq t_i\}} \frac{u_{ir{l}} - \xi_{c_{ir{l}}}}{s^2_{c_{ir{l}}}} \right),\, \sigma^2_{\alpha r\ell}
        \right\}
    \]
    where $\sigma^2_{\alpha r\ell} = 1 / (1/\sigma^2_\alpha + 1 / \sum_{{l}\in S^r_\ell} \sum_{\{i\mid {l}\leq t_i\}} s^2_{c_{ir{l}}})$.
\end{enumerate}
\end{algorithm}

\begin{algorithm}
\caption{MCMC step for $\pi_\beta$ and $\bm \beta_{1},\dots,\bm\beta_m$ based on the augmented likelihood. \label{alg:update_beta}}
\begin{enumerate}
    \item
    Recall from Web Appendix~\ref{ap:var_sel} that $S_l$ denotes a group of predictors.
    Let $B_r^{\star} = {\{l\mid \bm\beta_{rS_l}\ne \bm 0 \}}$ and $b_r^{\star} = |B_r|$ denote the indices and the number, respectively, of non-zero groups of coefficients in $\bm\beta_r$.
    Sample \[\pi_\beta\mid \bm\beta_1,\dots,\bm\beta_m \sim \mathrm{Beta}\left\{
        1 + \sum_{r=1}^m b_r^{\star},\,
        1 + \sum_{r=1}^m (L - b_r^{\star})
    \right\}\]
    where $L$ is the number of groups of predictors.
    \item
    For $r=1,\dots,m$, update $\bm \beta_{r}$ similarly to Section~2.5 of \citet{Held2006}:
    \begin{enumerate}
        \item
        Let $B_r = \{j\mid \beta_{rj}\ne 0\}$ and $b_r = |B_r|$ denote the indices and the number, respectively, of non-zero elements in $\bm\beta_r$.
        Also, let
        $\Sigma_{B_r} = \{I_{b_r}/\sigma^2_\beta + \sum_{i=1}^n \sum_{{l}=1}^{t_i} \bm x_{iB_r}\bm x_{iB_r}^\top / s_{c_{ir{l}}}^2\}^{-1}$ where $\bm x_{iB_r}$ is the $b_r$-dimensional subvector of $\bm x_i\ =(x_{i1},\dots,x_{ip})$ indexed by $B_r$,
        and\\
        $\bm\mu_{B_r} = \Sigma_{B_r} \sum_{i=1}^n \bm x_{iB_r} \sum_{{l}=1}^{t_i} (u_{ir{l}} - \xi_{c_{ir{l}}} - \alpha_{r{l}}) / s_{c_{ir{l}}}^2$.
        \item
        Update the set $B_r^{\star}$, i.e.\ which groups of predictors are included. For $l=1,\dots,L$:
        \begin{enumerate}
            \item
            Construct a Metropolis-Hastings proposal ${B_r^{\star}}'$ from $B_r^{\star}$ by adding $l$ to ${B_r^{\star}}'$ if $l\in B_r^{\star}$
            and removing $l$ from ${B_r^{\star}}'$ otherwise.
            Let ${b_r^{\star}}' = |{B_r^{\star}}'|$.
            Based on the proposal ${B_r^{\star}}'$, we define $B_r'$ and $b_r'$ analogously to $B_r$ and $b_r$.
            \item
            Set
            $B_r^{\star} = {B_r^{\star}}'$ with probability
            \[
                \min\left\{1,\,
                    \frac{
                        |\Sigma_{B_r^{'}}|^{1/2}\, \sigma_\beta^{b_r}\,
                        \exp( \bm\mu_{B_r'}^\top \Sigma_{B_r'}^{-1} \bm\mu_{B_r'} /2)\, \pi_\beta^{{b_r^{\star}}' - b_r^{\star}}
                    }{
                        |\Sigma_{B_r}|^{1/2}\, \sigma_\beta^{b_r'}\,
                        \exp( \bm\mu_{B_r}^\top \Sigma_{B_r}^{-1} \bm\mu_{B_r} /2)\, (1 - \pi_\beta)^{{b_r^{\star}}' - b_r^{\star}}
                    }
                \right\}
            \]
        \end{enumerate}
        \item
        Set $\beta_{rj}=0$ for $j\notin B_r$ and sample
        \[\bm\beta_{rB_r}\mid B_r,\bm u,\bm c,\bm\alpha_1,\dots,\bm\alpha_{{t_{\max}}} \sim\mathcal{N}(\mu_{B_r}, \Sigma_{B_r})\]
    \end{enumerate}
\end{enumerate}
\end{algorithm}

\subsection{Global MCMC from Bayesian nonparametrics}

In the data augmentation,
the latent variables $\bm u$ and component indicators $\bm c$ can strongly reflect the values for $\bm\alpha_{t}$
associated with
the current change point configuration.
Therefore, we also consider MCMC updates for the
change point indicators
$\gamma_t$ and $\bm z_t$ that directly use the unaugmented likelihood
in \eqref{eq:likelihood2}.
The updates are detailed in Algorithms~\ref{alg:update_gamma_unaugmented}
and \ref{alg:update_z_unaugmented}. They closely mirror Algorithms~\ref{alg:update_gamma_augmented} and \ref{alg:update_z_augmented}, respectively, which use the augmented likelihood.
The main difference is that values need to be proposed for $\bm\alpha_t$ as it is not integrated out from the likelihood
here.

\begin{breakablealgorithm}
\caption{Split-merge-shuffle step for $\gamma_t$ based on the unaugmented likelihood. \label{alg:update_gamma_unaugmented}}
\begin{enumerate}
    \item
    Perform a split or a merge step:
    \begin{enumerate}
        \item
        With probability $\mathds{1}[K=0] + \frac{1}{2}\mathds{1}[0<K<{t_{\max}}-1]$,
        perform a split step:
        \begin{enumerate}
            \item
            Sample a time interval ${\ell}'$ uniformly from $\{{\ell}\mid n_{\ell} \geq 2\}$.
            \item
            Split time interval ${\ell}'$ by setting $\gamma_{t'} = 1$ for a $t'$ sampled uniformly from $\{2 + \sum_{\kappa = 1}^{{\ell}'-1} n_\kappa,\,3 + \sum_{\kappa = 1}^{{\ell}'-1} n_\kappa,\,\dots,\, \sum_{\kappa = 1}^{{\ell}'} n_\kappa\}$.
            \item
            Sample corresponding cause-specific change points $\bm z_{t'}'$
            from the prior $p(\bm z_{t'}\mid \gamma_{t'})$.
            \item
            For each newly introduced cause-specific change point, propose values $\alpha_{r,\ell'+1}^{\star\prime}$ for $t$ corresponding to the newly introduced time interval by sampling from $\mathcal{N}(\alpha_{r,\ell'}^{\star},\, 1)$.
            Let $\bm\theta'$ denote the resulting parameter where the other change points, the other $\bm\alpha_t$ and $\bm\beta_{r}$ are left unchanged.
            \item
            Set $\bm\theta=\bm\theta'$ with probability
            \[
                \min\left\{1,\, \frac{2^{\mathds{1}[K > 0]}\, |\{{\ell}\mid n_{\ell} \geq 2\}|\, (n_{{\ell}'} - 1)\,p(\bm\theta')\, p(\bm y\mid \bm\theta')}{2^{\mathds{1}[K < {t_{\max}} - 2]}\, (K + 1)\,p(\bm z'_{t'}\mid \gamma_{t'})\,p(\bm\theta)\, p(\bm y\mid \bm\theta)
                \prod_{\{r\mid z_{rt'}' = 1 \}} \mathcal{N}(\alpha_{r,\ell'+1}^{\star\prime}\mid \alpha_{r,\ell'}^{\star},\, 1)
                }\right\}
            \]
        \end{enumerate}
        \item
        Otherwise, perform a merge step:
        \begin{enumerate}
            \item
            Sample a time interval ${\ell}'$ uniformly from $\{1,\dots,K\}$.
            \item
            Let $\bm\theta'$ denote the parameter resulting from merging time intervals ${\ell}'$ and $({\ell}'+1)$, and the corresponding removal of change points.
            Here, the baseline hazards $\alpha_{r\ell'}^\star$ from time interval ${\ell}'$ are used for the merged time interval while the $\alpha_{r,\ell'+1}^\star$ from time interval $({\ell}'+1)$ are discarded.
            \item
            Let $t_\text{old} = 1 + \sum_{\kappa=1}^{{\ell}'} n_\kappa$.
            Set $\bm\theta=\bm\theta'$ with probability
            \[
                \min\left\{1,\, \frac{
                2^{\mathds{1}[K < {t_{\max}} - 1]}\, K\,p(\bm z_{t_\textnormal{old}}\mid \gamma_{t_\textnormal{old}})\,p(\bm\theta')\, p(\bm y\mid \bm \theta')
                \prod_{\{r\mid z_{rt_\text{old}} = 1 \}} \mathcal{N}(\alpha_{r,\ell'+1}^\star\mid \alpha_{r\ell'}^\star,\, 1)
                }{
                2^{\mathds{1}[K > 1]}\, |\{{\ell}\mid n_{\ell}' \geq 2\}|\, (n_{{\ell}'}' - 1)\,p(\bm\theta)\, p(\bm y\mid \bm \theta)}\right\}
            \]
            where $n_{{\ell}'}'$ is the duration of the merged time interval.
        \end{enumerate}
    \end{enumerate}
    \item
    If $K > 0$, perform a shuffle step:
    \begin{enumerate}
        \item
        Sample a time interval ${\ell}'$ uniformly from $\{1,\dots,K\}$. Let $t_\text{old}=1 + \sum_{\kappa=1}^{{\ell}'} n_\kappa$.
        \item
        Sample a new overall change point $t'$ uniformly from\\
        $\{t_\text{old} - n_{{\ell}'},\,{t_\text{old} - n_{{\ell}'}+1},\,\dots,\, {t_\text{old} + n_{{\ell}'+1} - 1}\}$.
        \item
        Let $\bm\theta'$ denote the parameter resulting from setting
        $\gamma'_{t'}=1$, $\bm z'_{t'} = \bm z_{t_\text{old}}$ and,
        if $t'\ne t_\text{old}$,
        $\gamma'_{t_\text{old}} = 0$
        and
        $\bm z_{t_\text{old}} = \bm 0$,
        while leaving the baseline hazards $\alpha_{r\ell}^\star$ corresponding to each time interval the same. That is, only the duration of the time intervals change.
        \item
        Set $\bm\theta=\bm\theta'$ with probability
        \[
            \min\left\{1,\, \frac{
           p(\bm\theta')\, p(\bm y\mid \bm \theta')
            }{
            p(\bm\theta)\, p(\bm y\mid \bm \theta)}\right\}
        \]
    \end{enumerate}
\end{enumerate}
\end{breakablealgorithm}

\begin{algorithm}
\caption{Metropolis-Hastings step for $\bm z_{t}$ based on the unaugmented likelihood. \label{alg:update_z_unaugmented}}
If $K>0$, update $\bm z_t$ corresponding to a randomly sampled time interval:
\begin{enumerate}
    \item
    Sample ${\ell}'$ uniformly from $\{2,\dots,K+1\}$. Let $t' = 1 + \sum_{\kappa=1}^{{\ell}' - 1} n_{\ell}$.
    \item
    Sample cause-specific change points $\bm z_{t'}'$
    from the prior $p(\bm z_{t'}\mid \gamma_{t'})$.
    \item
    For each newly introduced cause-specific change point (i.e.\ $z'_{rt'} = 1$ and $z_{rt'} = 0$), propose values $\alpha_{r\ell'}^{\star\prime}$ corresponding to the new cause-specific time interval that it starts by sampling from $\mathcal{N}(\alpha_{r,\ell'-1}^{\star},\, 1)$.
    For each removed cause-specific change point (i.e.\ $z_{rt'} = 1$ and $z'_{rt'} = 0$), carry forward the baseline hazards $\alpha_{r,\ell'-1}^{\star}$ from time interval $({\ell}'-1)$ to time interval ${\ell}'$.
    Let $\bm\theta'$ denote the resulting parameter where the other change points, the other $\alpha_{r\ell}^\star$ and $\bm\beta_{r}$ are left unchanged.
    \item
    Set $\bm\theta=\bm\theta'$ with probability
    \[
        \min\left\{1,\, \frac{p(\bm z_{t'}\mid \gamma_{t'})\, p(\bm\theta')\, p(\bm y\mid \bm\theta')
        \prod_{\{r\mid z_{rt'} = 1\, \&\, z_{rt'}' = 0 \}} \mathcal{N}(\alpha_{r\ell'}^{\star}\mid \alpha_{r,\ell-1}^{\star},\, 1)
        }{p(\bm z'_{t'}\mid \gamma_{t'})\, p(\bm\theta)\, p(\bm y\mid \bm\theta)
        \prod_{\{r\mid z_{rt'}' = 1\, \&\, z_{rt'} = 0 \}} \mathcal{N}(\alpha_{r\ell'}^{\star\prime}\mid \alpha_{r,\ell'-1}^{\star},\, 1)
        }\right\}
    \]
\end{enumerate}
\end{algorithm}

We are now ready to state the full MCMC step used for the Multivariate Bernoulli detector in Algorithm~\ref{alg:mvb_detector}.
The computationally most expensive part is typically Step~\ref{step:augmentation} which samples the augmented data using Algorithm~\ref{alg:augmentation}.
More precisely,
Step~\ref{step:sample_c} of Algorithm~\ref{alg:augmentation}
is most expensive.
Fortunately, it is embarrassingly parallel across individuals $i$ and times ${l}$ such that its computation can be readily sped up using multicore computing.
In the global step,
the unaugmented likelihood $p(\bm y\mid \bm\theta)$ in \eqref{eq:likelihood2}
can also be expensive to compute.
See Section~4.1.3 and Appendix~A of \citet{King2021} for ways to speed up evaluation of such a likelihood.

\begin{algorithm}
\caption{Local-global MCMC step for the Multivariate Bernoulli detector. \label{alg:mvb_detector}}
\begin{enumerate}
    \item
    Local step:
    \begin{enumerate}
        \item \label{step:augmentation}
        Sample the data augmentation $\bm u$ and $\bm c$ per Algorithm~\ref{alg:augmentation}.
        \item
        Update $\gamma_t$ and $\bm z_t$ using the augmented likelihood per Algorithm~\ref{alg:update_gamma_augmented}.
        \item
        Update $\bm z_t$ using the augmented likelihood per Algorithm~\ref{alg:update_z_augmented}.
        \item
        Update $\bm\alpha_t$ using the augmented likelihood per Algorithm~\ref{alg:update_alpha}.
        \item
        Update $\bm\beta_1,\dots,\bm\beta_m$ using the augmented likelihood per Algorithm~\ref{alg:update_beta}.
    \end{enumerate}
    \item
    Global step:
    \begin{enumerate}
        \item
        Update $\gamma_t$, $\bm z_t$ and $\bm\alpha_t$ using the unaugmented likelihood $p(\bm y\mid\bm\theta)$ per Algorithm~\ref{alg:update_gamma_unaugmented}.
        \item
        Update $\bm z_t$ and $\bm\alpha_t$ using the unaugmented likelihood $p(\bm y\mid\bm\theta)$ per Algorithm~\ref{alg:update_z_unaugmented}.
    \end{enumerate}
\end{enumerate}
\end{algorithm}

\FloatBarrier

\section{Simulation study}
\label{ap:simul}

Here we present a comprehensive simulation study, which includes a wide variety of scenarios, replicate simulated data sets, and comparisons with maximum likelihood estimation and the model by \citet{King2021}.

\subsection{Simulation study without change points}
\label{sec:simul_null}

We apply the Multivariate Bernoulli detector to data simulated without change points.
We simulate $n=100$ observations from the model in Section~\ref{sec:model} with $m=3$ risks, no predictors ($p=0$), no censoring ($\delta_i = 1$), and constant baseline hazard $\bm\alpha_t = (-2,-3,-4)$, $t\geq 1$.
We fit our model with ${t_{\max}} = \max_i t_i = 26$.
We run the MCMC chain for 100000 iterations, discarding the first 10000 as burn-in.

The MCMC estimate of the posterior probability of no change points is $0.83$. Also, the estimate of the Bayes factor of no change points from Web Appendix~\ref{ap:BF} is 1.89 which is in line with the absence of change points.
Finally, the 95\% credible intervals for $\bm \alpha_t$
are from $(-2.2, -3.2, -10.4)$ to $(-1.9, -2.7, -3.4)$ which include the true values $\bm\alpha_t = (-2,-3,-4)$.

\subsection{Simulation study with change points}
\label{sec:simul}

Here, focus is on inference on change points.
We simulate $n=300$ observations from the model in Section~\ref{sec:model} of the main manuscript
with
$t_{\max}=20$,
$m=3$ risks, no covariates ($p=0$) and baseline hazard
with two change points: one at $t=6$ involving all three risks
and another at $t=13$ affecting only risks 1 and 2.
Specifically,
$\bm\alpha_{t} = (-9, -9, -9)$ before the first change point ($t=1,\dots,5$),
$\bm\alpha_{t} = (-4, -3, -3)$ between the two change points ($t=6,\dots,12$)
and
$\bm\alpha_{t} = (-2, -2, -3)$ after the second change point ($t=13,\dots$).
We consider three scenarios with (i) no censoring beyond those with $T_i=t_{\max} + 1$, and (ii) 10\% and (iii) 50\% of the individuals being censored, which are selected at random.
For those censored individuals,
we sample the censoring time $C_i$ uniformly
from the set
$\{1,\dots,T_i - 1\}$.
We fit our model using 100000 MCMC iterations, discarding the first 10000 as burn-in which takes around four minutes with an AMD Ryzen 7950X CPU with a base clock speed of 4.5 GHz.

\begin{figure}[tbp]
\centering
\includegraphics[width=\textwidth]{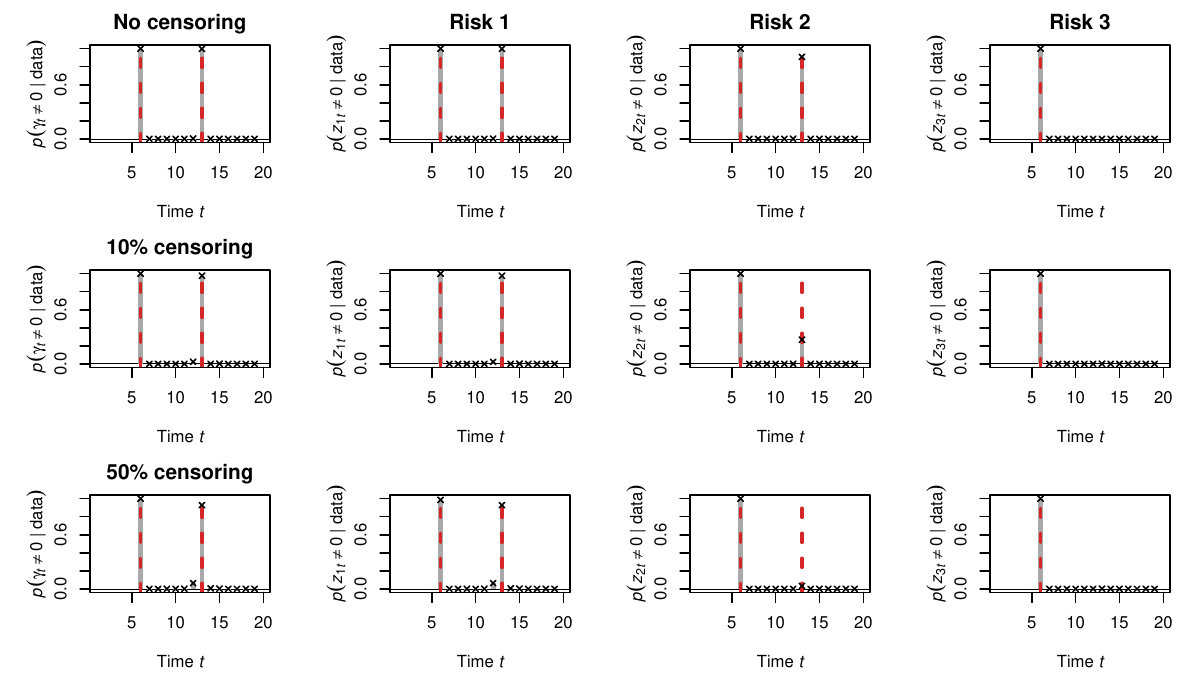}
\caption{Simulation study: Posterior probabilities for the presence of a change point for the overall (left column) and cause-specific (other columns) hazard functions.
The gray lines correspond to posterior probabilities.
The dashed red lines are drawn in correspondence of the true change points.}
\label{fig:simul_cp}
\end{figure}

\begin{figure}
\centering
\includegraphics[width=\textwidth]{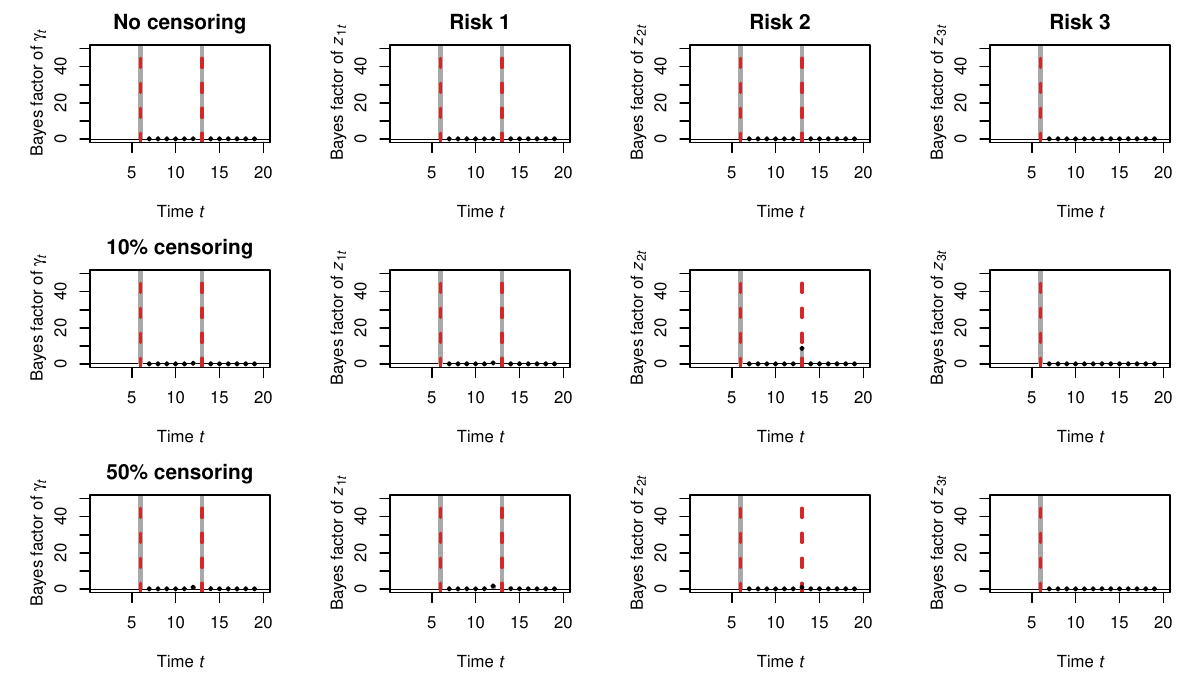}
\caption{Simulation study: Bayes factors for the presence of a change point for the overall (left column) and cause-specific (other columns) hazard functions.
The gray lines correspond to Bayes factors, some of which are outside the plotting range.
The dashed red lines are drawn in correspondence of the true change points.
\label{fig:simul_cp_BF}}
\end{figure}

\begin{figure}
\centering
\includegraphics[width=\textwidth]{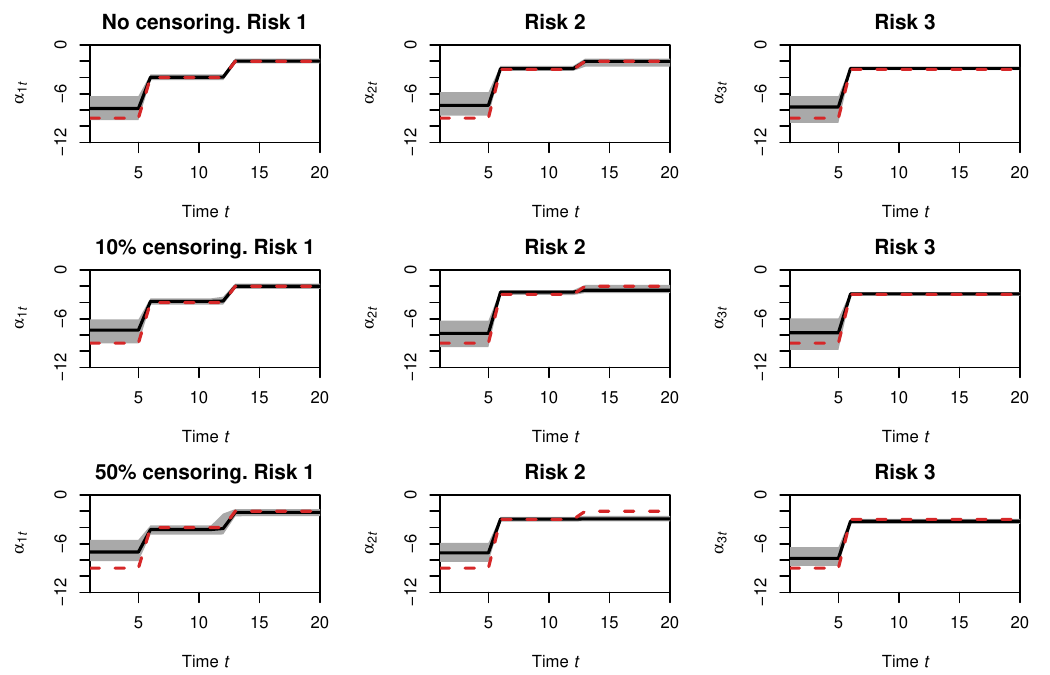}
\caption{Simulation study: Posterior inference on the baseline hazard parameter $\alpha_{rt}$ from the Multivariate Bernoulli detector. Solid lines represent posterior means and shaded areas correspond to 95\% credible intervals.
The dashed lines are drawn in correspondence of the true $\alpha_{rt}$.}
\label{fig:simul_alpha}
\end{figure}

We summarize the resulting inference on change points in Web Figure~\ref{fig:simul_cp} using the
posterior probabilities of $\gamma_t=1$
and $z_{rt}=1$ (see Web Figure~\ref{fig:simul_cp_BF} for corresponding Bayes factors).
The values are
shown only for those times $t\in \mathcal{T}$ at which change points are allowed.
Our model recovers both the overall and cause-specific change points with high accuracy in the absence of censoring.
With 10\% censoring, inference on the relatively minor second change point in the second risk is more uncertain.
This cause-specific change point is missed with 50\% censoring.
Inference on $\alpha_{rt}$,
presented in Web Figure~\ref{fig:simul_alpha}, similarly shows accurate recovery across all risks only for scenarios~(i) and (ii) with limited censoring.
Finally, the estimate of the Bayes factor in Web Appendix~\ref{ap:BF} and the posterior probability of no change points are
zero in all three scenarios.

\begin{figure}
\centering
\includegraphics[width=\textwidth]{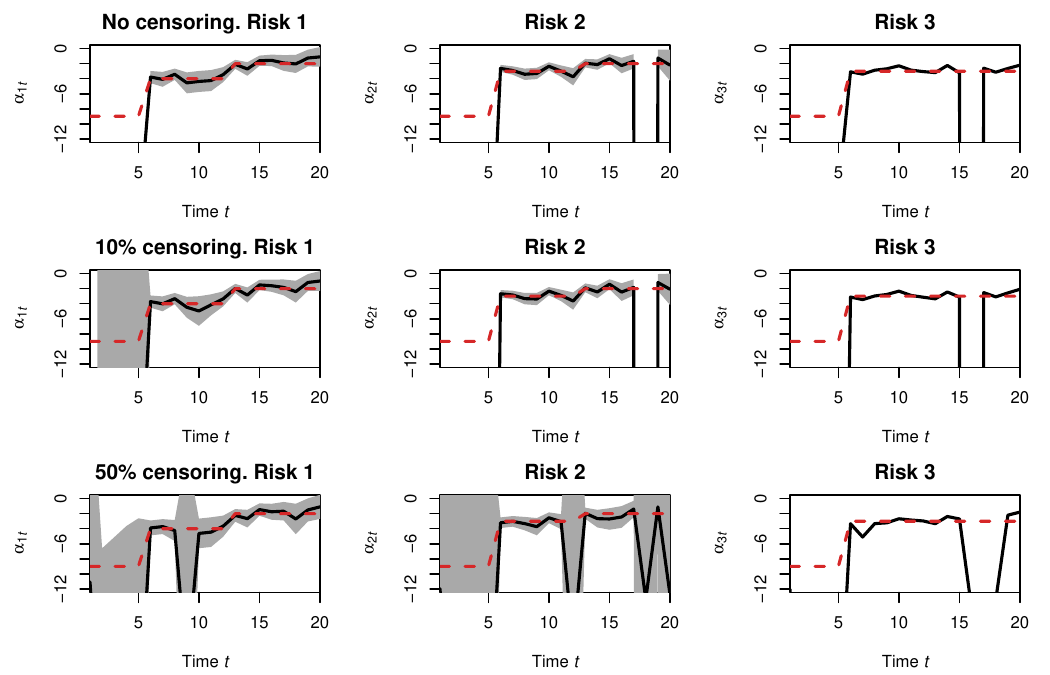}
\caption{Simulation study:  Maximum likelihood estimates of the baseline hazard parameter $\alpha_{rt}$ (solid lines) with their 95\% confidence intervals demarcated by shaded areas.
The dashed lines are drawn in correspondence of the true $\alpha_{rt}$.
The confidence interval is not available for some $\alpha_{rt}$ due to numerical issues.}
\label{fig:simul_nnet}
\end{figure}

\begin{figure}
\centering
\includegraphics[width=\textwidth]{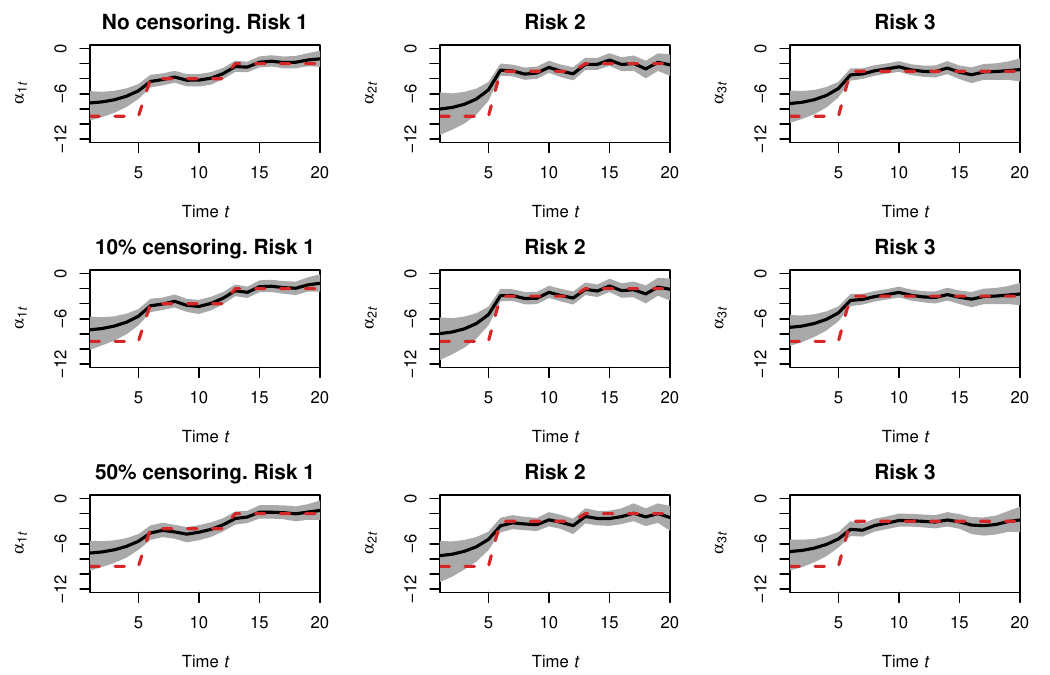}
\caption{Simulation study: Posterior inference on the baseline hazard parameter $\alpha_{rt}$ from the model by \citet{King2021}. Solid lines represent posterior means and shaded areas correspond to 95\% credible intervals.
The dashed lines are drawn in correspondence of the true $\alpha_{rt}$.}
\label{fig:simul_brea}
\end{figure}

We also fit the models described in Section~\ref{sec:compare} 
with results shown in Web Figures~\ref{fig:simul_nnet} and \ref{fig:simul_brea}.
The maximum likelihood estimates are non-smooth and can be far from the true values of $\alpha_{rt}$, showing the need for regularization of such estimation.
The model by \citet{King2021} provides smoother estimates, though with generally wider confidence intervals than the Multivariate Bernoulli detector.
A reason for this might be that the Multivariate Bernoulli detector can detect and exploit sharing of information across risks while \citet{King2021} smooth each risk independently.

\FloatBarrier
\subsection{Simulation study with replicates}
\label{sec:simul_rep}

\subsubsection{Setup}

\begin{table}
    \centering
\begin{threeparttable}
    \caption{Simulation scenarios.}
    \label{tab:scenarios}
    \vspace{0.3cm}
    \begin{tabular}{l|llll}
    \textbf{Scenario} & \textbf{Censoring level} & $n$ & $p$ & $m$ \\
    \hline
    \multicolumn{5}{l}{\emph{Varying censoring level}} \\
    i & 0\% & 500 & 3 & 2 \\
    ii & 10\% & 500 & 3 & 2 \\
    iii & 50\% & 500 & 3 & 2 \\
   \hline
    \multicolumn{5}{l}{\emph{Varying number of observations}} \\
    iv & 10\% & 200 & 3 & 2 \\
    ii & 10\% & 500 & 3 & 2 \\
    v & 10\% & 1000 & 3 & 2 \\
   \hline
    \multicolumn{5}{l}{\emph{Varying number of predictors}} \\
    ii & 10\% & 500 & 3 & 2 \\
    vi & 10\% & 500 & 5 & 2 \\
    \hline
    \multicolumn{5}{l}{\emph{Varying number of competing risks with $n=500$}} \\
    ii & 10\% & 500 & 3 & 2 \\
    vii & 10\% & 500 & 3 & 3 \\
    viii & 10\% & 500 & 3 & 4 \\
    \hline
    \multicolumn{5}{l}{\emph{Varying number of competing risks with $n=1000$}} \\
    v & 10\% & 1000 & 3 & 2 \\
    ix & 10\% & 1000 & 3 & 3 \\
    x & 10\% & 1000 & 3 & 4 \\
    \hline
    \end{tabular}
    \vspace{0.3cm}
    \begin{tablenotes}
        \item In the table, $n$ denotes the number of observations, $p$ the number of predictors and $m$ the number of competing risks.
    \end{tablenotes}
\end{threeparttable}
\end{table}

We consider ten scenarios by varying
(i) censoring level;
(ii) number of observations $n$;
(iii) number of predictors $p$;
(iv) number of competing risks $m$.
The scenarios are detailed in Web Table~\ref{tab:scenarios}.
In all scenarios, $t_{\max} = 15$ and the baseline hazard has two change points:
the first change point is at $t=6$ and involves all risks except in scenarios with $m=4$ risks when the fourth risk does not have a change point at $t=6$.
The second change point is at $t=13$ and involves only the first and third risks.
Specifically, for scenarios with $m=4$,\\
$\bm\alpha_t = (-5,-5,-5,-4)$ before the first change point ($t=1,\dots,5$),\\
$\bm\alpha_t = (-4,-3,-3,-4)$ between the two change points ($t=6,\dots,12$)
and\\
$\bm\alpha_t = (-2,-3,-2,-4)$ after the second change point ($t=13,\dots$).
For scenarios with $m=2,3$ risks,
the baseline hazard $\bm\alpha_t$ 
includes the first two or three elements of the previous vectors.
Regression coefficients $\beta_{rj}$ and (categorical) predictors $x_{rj}$ are sampled independently and uniformly from the sets $\{-0.1, 0, 0.1\}$ and $\{0,1\}$, respectively.
Then, data are simulated as in Section~\ref{sec:simul}.
We replicate each scenario 128 times.

For each of the 128 replicate data sets,
we fit the Multivariate Bernoulli detector as described in Section~\ref{sec:simul}.
For comparison, we also estimate $\bm\alpha_t$ and $\beta_{rj}$ by maximum likelihood, as described in Section~\ref{sec:mle},
and fitting the model by \citet{King2021}, as described in Section~\ref{sec:King2021}.

\subsubsection{Results}

To summarize inference across scenarios and replicates,
we focus on mean squared error (MSE) and bias of the estimates for the baseline hazard $\bm\alpha_t$, the regression coefficients $\beta_{rj}$ and the number of change points $K$.
In addition to $K=\sum_{t\in\mathcal{T}}\gamma_t$, we also consider the number of cause-specific change points $K_r = \sum_{t\in\mathcal{T}}z_{rt}$ where  $z_{rt} = \mathds{1}[\alpha_{rt} \ne \alpha_{r(t-1)}]$.
The MSE and bias are averaged across time, risks and predictors.
In more details, let us denote the estimates of the parameters by (i) $\widehat{\bm\alpha}_t$; (ii) $\widehat{\beta}_{rj}$; (iii) $\widehat{K}$; (iv) $\widehat{K}_r$. Here, the estimates are either the posterior mean or the maximum likelihood estimate.
We compute, for each replicate,
\begin{align*}
    \textnormal{MSE}_\alpha &= \frac{1}{m\, t_{\max}} \sum_t\sum_r (\widehat{\alpha}_{rt} - \alpha_{rt})^2 \\
    \textnormal{bias}_\alpha &= \frac{1}{m\, t_{\max}} \sum_t\sum_r (\widehat{\alpha}_{rt} - \alpha_{rt}) \\
    \textnormal{MSE}_\beta &= \frac{1}{m\, p} \sum_j\sum_r (\widehat{\beta}_{rj} - \alpha_{rt})^2 \\
    \textnormal{bias}_\beta &= \frac{1}{m\, p} \sum_j\sum_r (\widehat{\beta}_{rj} - \alpha_{rt}) \\
    \textnormal{MSE}_K &= (\widehat{K} - K)^2 \\
    \textnormal{bias}_K &= \widehat{K} - K \\
    \textnormal{MSE}_{K_r} &= \frac{1}{m} \sum_r (\widehat{K}_r - K_r)^2 \\
    \textnormal{bias}_{K_r} &= \frac{1}{m} \sum_r (\widehat{K}_r - K_r) \\
\end{align*}

\begin{figure}
\centering
\includegraphics[width=\textwidth]{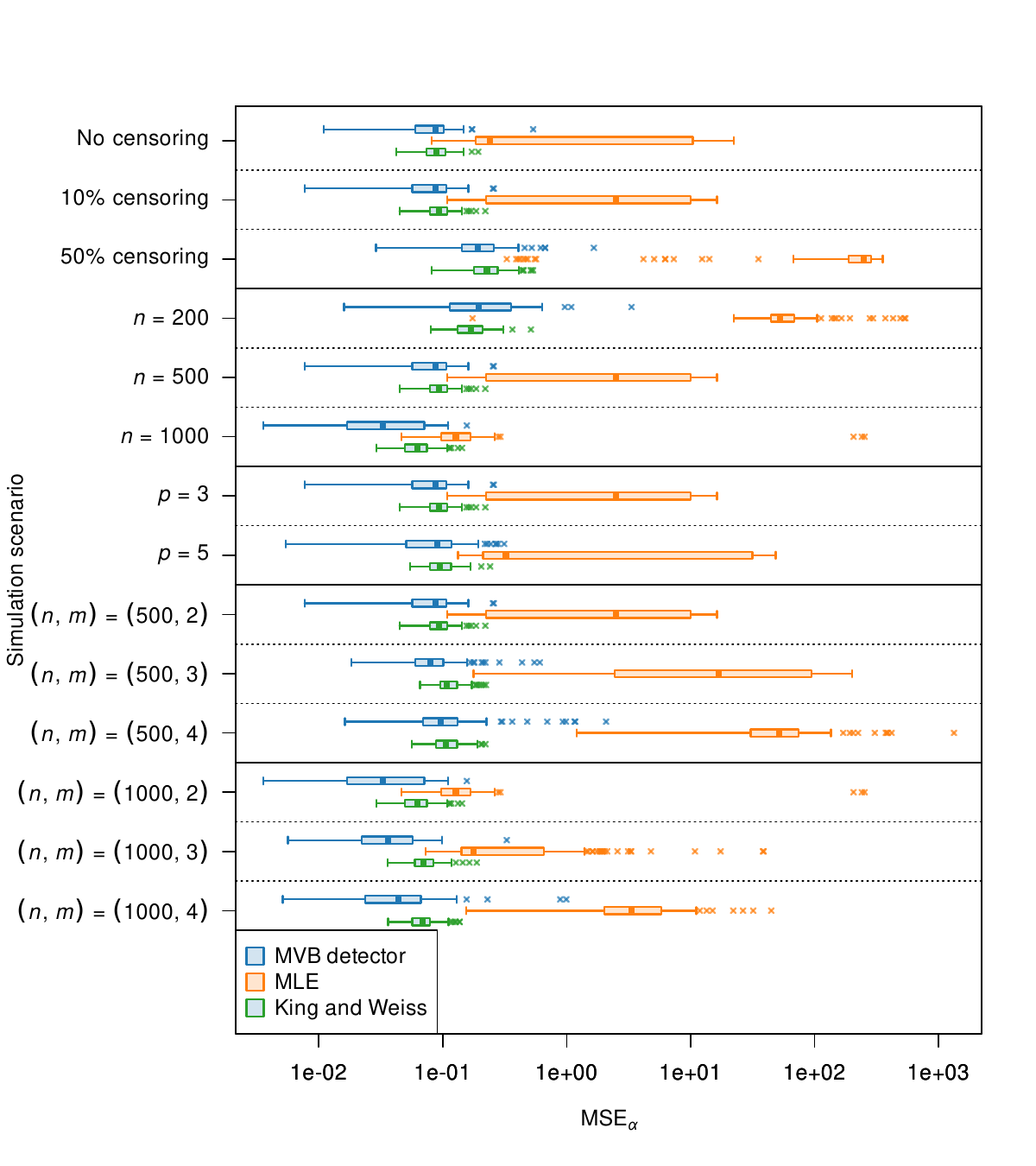}
\caption{Simulation study: For each scenario, we plot the distribution of the mean squared error (MSE) of the inference on the baseline hazard parameter $\alpha_{rt}$ across the 128 replicates from the Multivariate Bernoulli (MVB) detector, maximum likelihood estimation (MLE) and the model by \citet{King2021}.}
\label{fig:simul_MSE_alpha}
\end{figure}

\begin{figure}
\centering
\includegraphics[width=\textwidth]{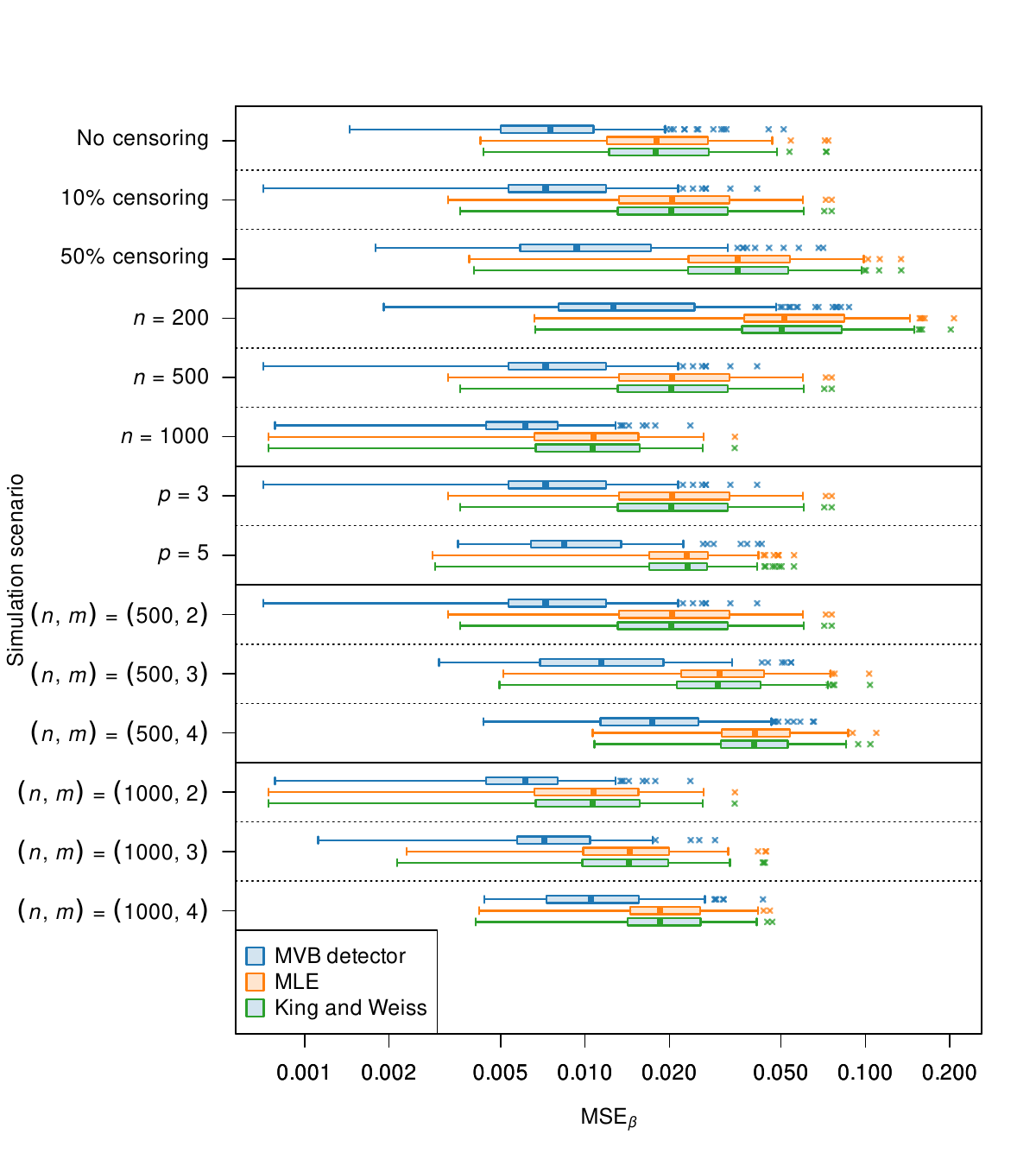}
\caption{Simulation study: For each scenario, we plot the distribution of the mean squared error (MSE) of the inference on the regression coefficients $\beta_{rj}$ across the 128 replicates from the Multivariate Bernoulli (MVB) detector, maximum likelihood estimation (MLE) and the model by \citet{King2021}.}
\label{fig:simul_MSE_beta}
\end{figure}

\begin{figure}
\centering
\includegraphics[width=\textwidth]{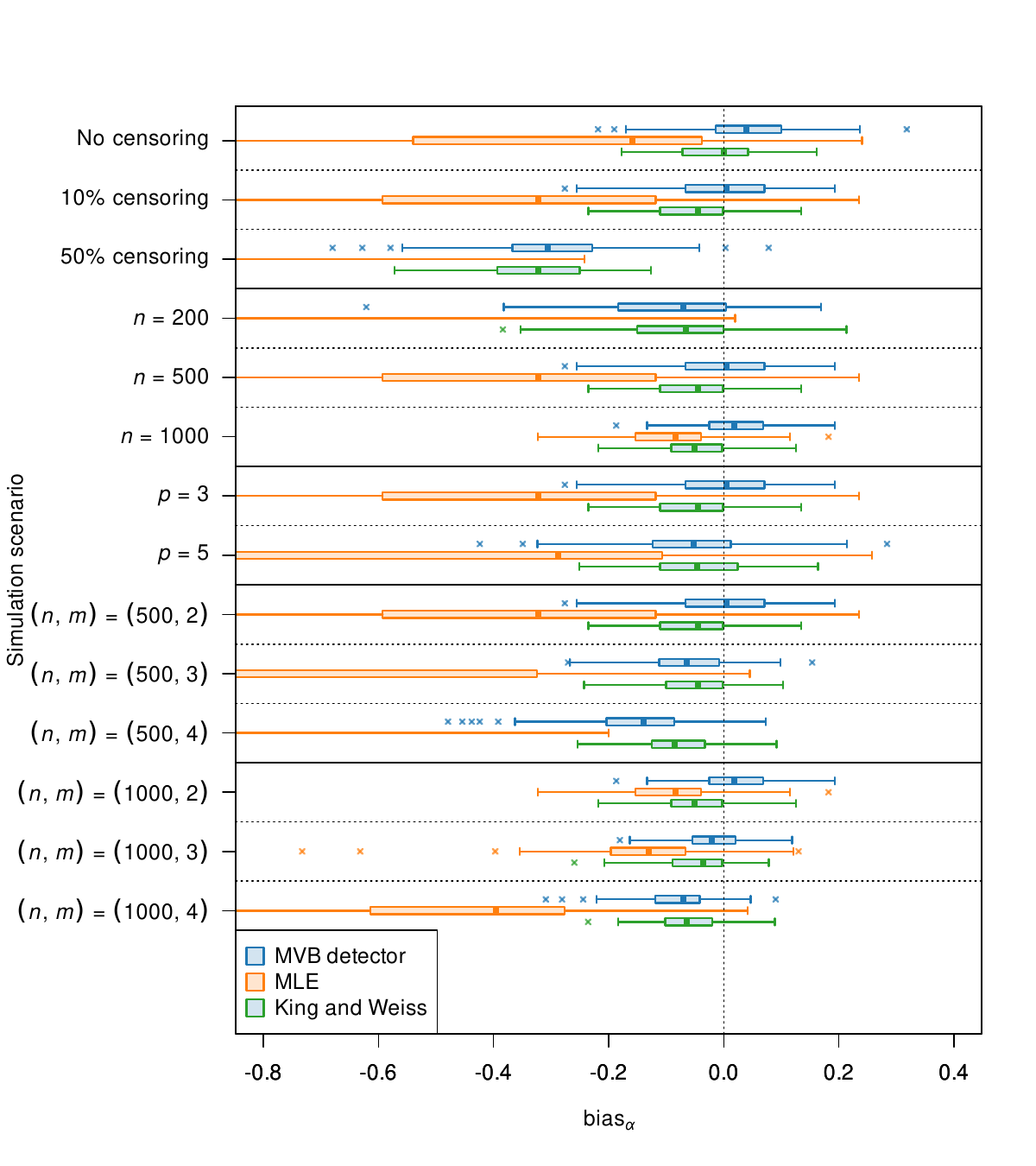}
\caption{Simulation study: For each scenario, we plot the distribution of the average bias of the estimates of the baseline hazard parameter $\alpha_{rt}$ across the 128 replicates from the Multivariate Bernoulli (MVB) detector, maximum likelihood estimation (MLE) and the model by \citet{King2021}. Some results for the MLE are outside the plotting range.}
\label{fig:simul_bias_alpha}
\end{figure}

\begin{figure}
\centering
\includegraphics[width=\textwidth]{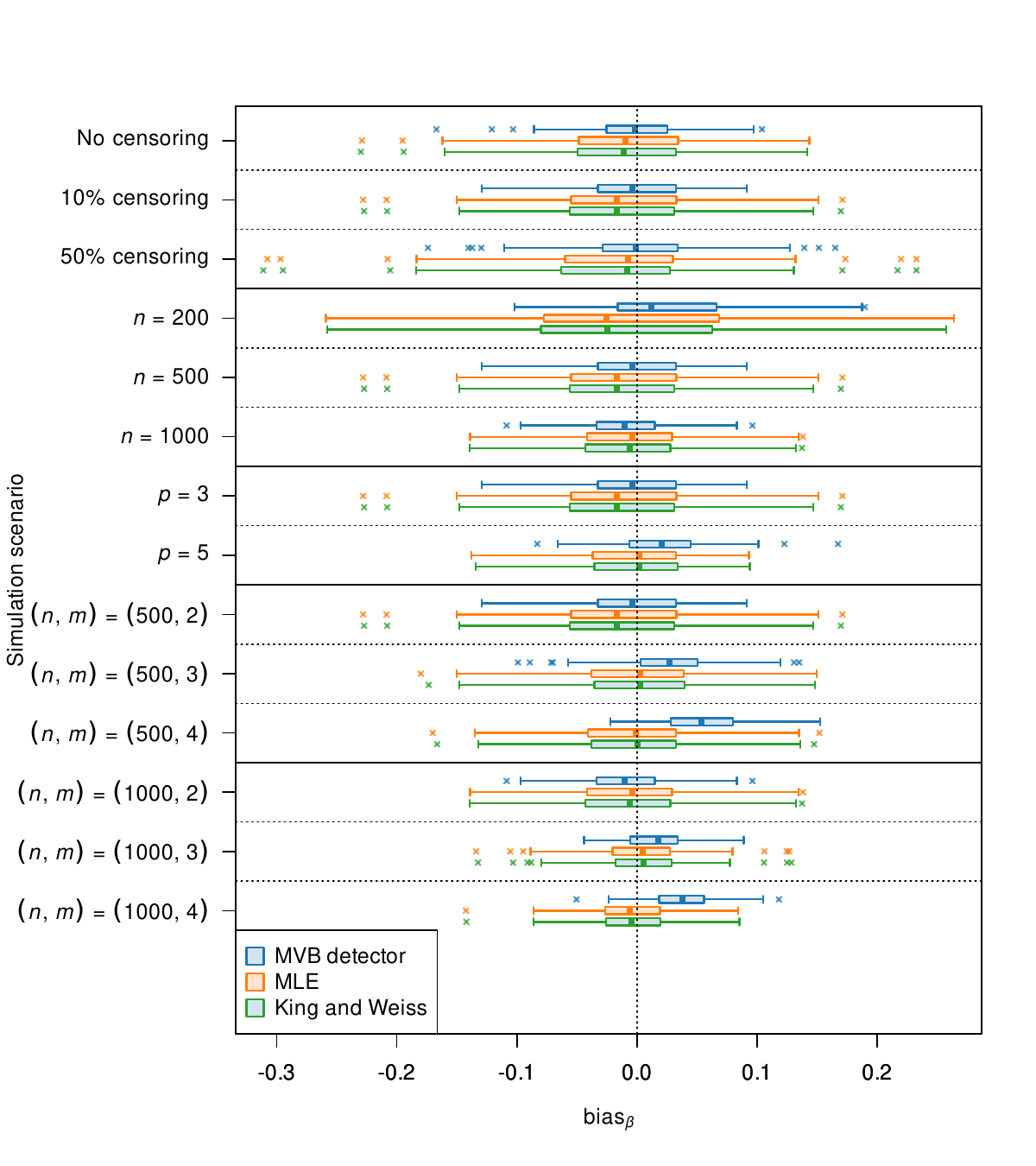}
\caption{Simulation study: For each scenario, we plot the distribution of the average bias of the inference on the regression coefficients $\beta_{rj}$ across the 128 replicates from the Multivariate Bernoulli (MVB) detector, maximum likelihood estimation (MLE) and the model by \citet{King2021}.}
\label{fig:simul_bias_beta}
\end{figure}

\begin{figure}
\centering
\includegraphics[width=\textwidth]{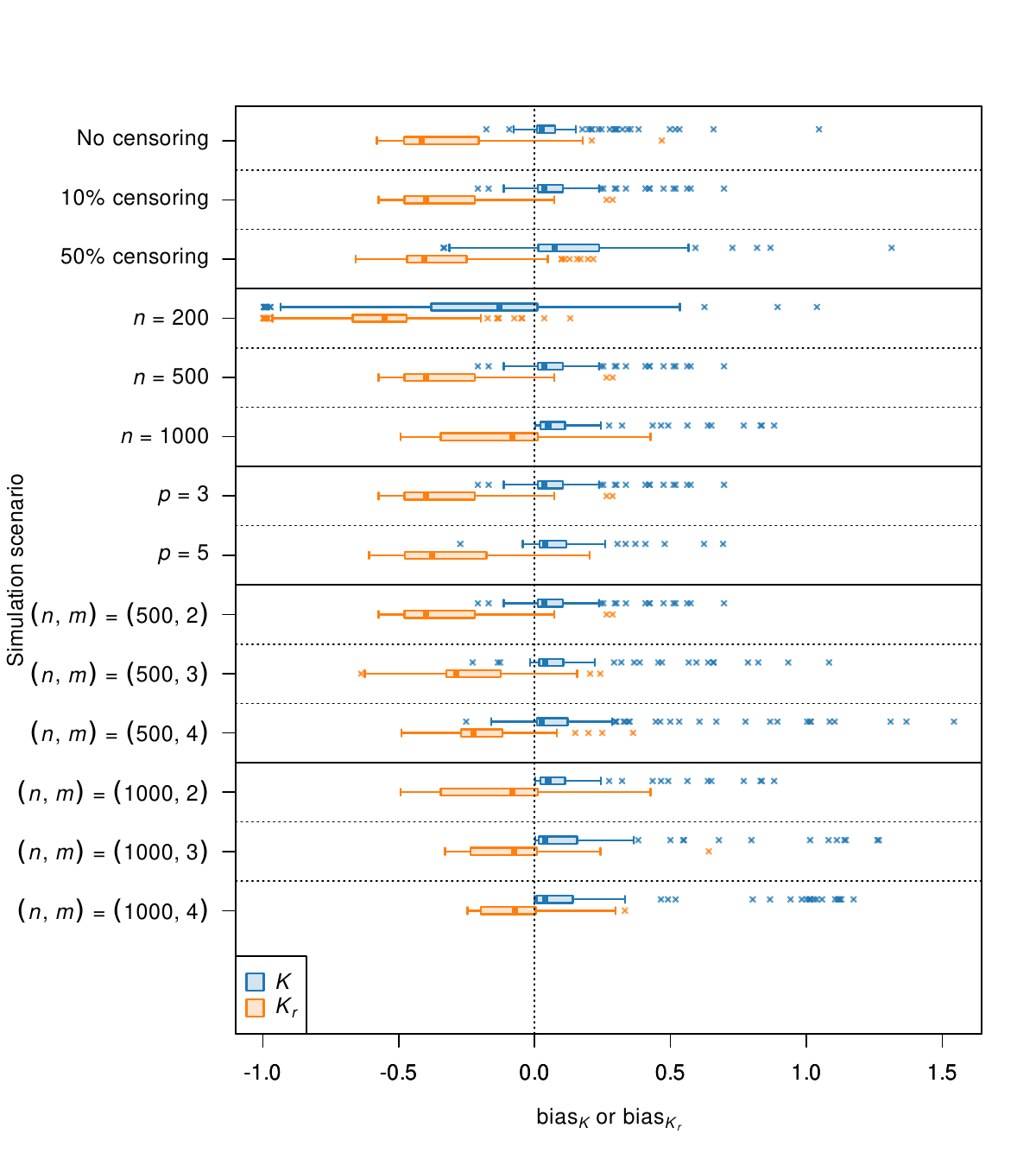}
\caption{Simulation study: For each scenario, we plot the distribution of the average bias of the inference on the number of change points $K$ and cause-specific change points $K_r$ across the 128 replicates from the Multivariate Bernoulli detector.}
\label{fig:simul_bias_K}
\end{figure}

\begin{figure}
\centering
\includegraphics[width=\textwidth]{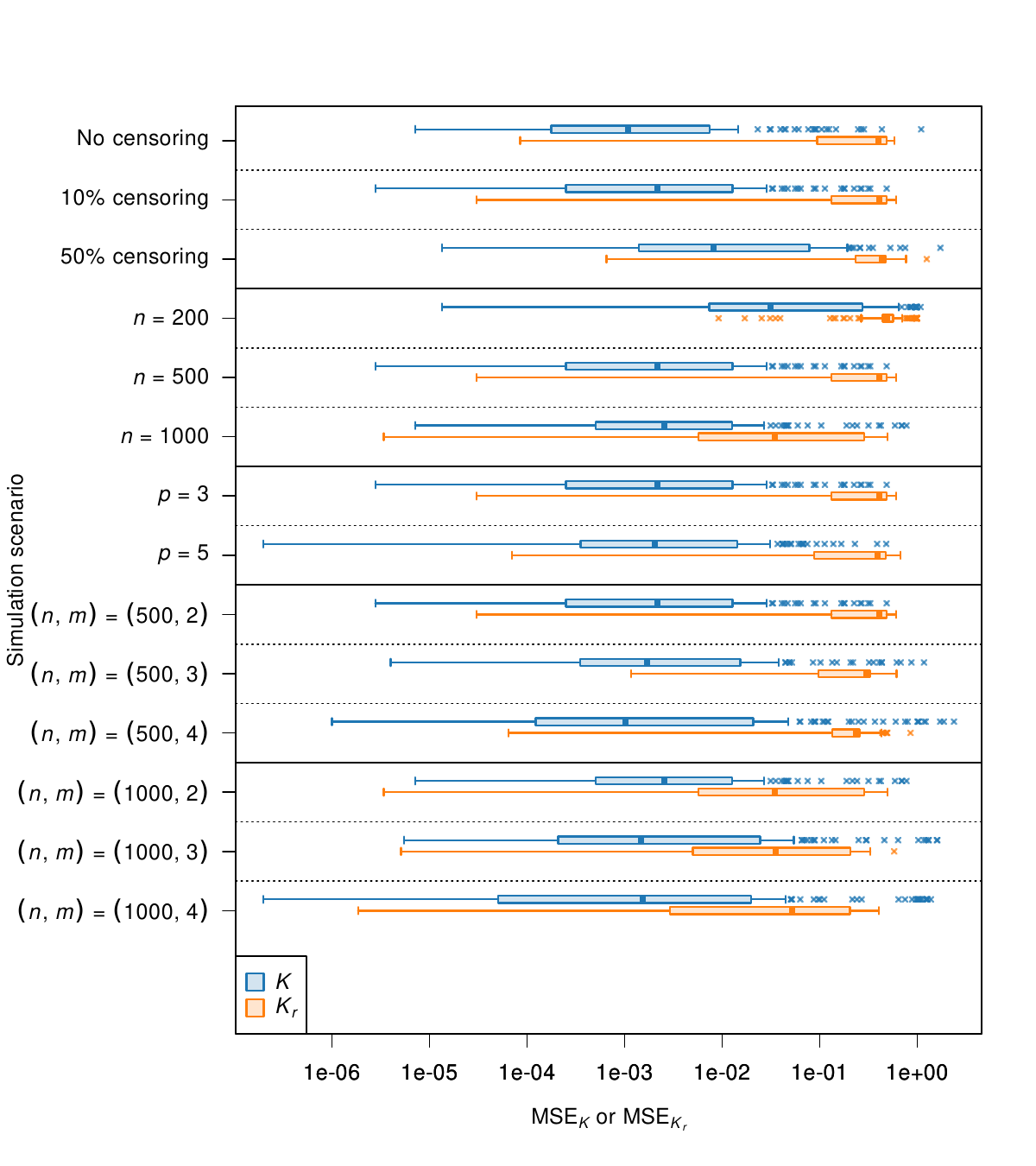}
\caption{Simulation study: For each scenario, we plot the distribution of the mean squared error (MSE) of the inference on the number of change points $K$ and cause-specific change points $K_r$ across the 128 replicates from the Multivariate Bernoulli detector.}
\label{fig:simul_MSE_K}
\end{figure}

The results are summarized in
Web Figures~\ref{fig:simul_MSE_alpha} through \ref{fig:simul_MSE_K}.
Note that some scenarios are visualised multiple times in the figures for ease of comparison. 
In terms of mean squared error, the Multivariate Bernoulli detector outperforms both maximum likelihood estimation and the model by \citet{King2021} in almost all scenarios (see Web Figures~\ref{fig:simul_MSE_alpha} and \ref{fig:simul_MSE_beta}).
Maximum likelihood estimation presents large $\textnormal{MSE}_\alpha$,
suggesting unstable estimation for the unconstrained approach: we discuss how unconstrained estimation can lead to unstable estimation in Section~\ref{sec:intro} of the main manuscript.

In terms of bias (see Web Figures~\ref{fig:simul_bias_alpha} and \ref{fig:simul_bias_beta}),
the comparison is more nuanced.
For the baseline hazards,
maximum likelihood estimation performs the worst with significant underestimation of $\bm\alpha_t$.
On the other hand, the Multivariate Bernoulli detector and the model by \citet{King2021} perform similarly in terms of $\textnormal{bias}_\alpha$.
For the regression coefficients,
the Multivariate Bernoulli detector overestimates $\beta_{rj}$ on average for some scenarios where the other two methods perform well, which is likely a result of underestimation of the baseline hazards due to prior shrinkage as shown in Web Figure~\ref{fig:simul_bias_alpha}.
We note that, nonetheless, the Multivariate Bernoulli detector performs best in terms of mean squared error as shown in Web Figure~\ref{fig:simul_MSE_beta}.
The results suggest that the shrinkage induced by the prior specification in the Multivariate Bernoulli detector increases bias but reduces variance even more, resulting in lower error compared to the other two methods.
This is
typical of shrinkage methods such as the James-Stein estimator and Bayesian shrinkage priors in general \citep[e.g.][]{Polson2019}.

Inference on the number of change points
is presented in Web Figure~\ref{fig:simul_bias_K}.
Recall that the MLE and the model by \citet{King2021} do not estimate number of change points.
The number of cause-specific change points $K_r$ is typically underestimated while the number of overall change points $K$ is overestimated by our approach. 
This is unsurprising since
multiple cause-specific change points constitute a single overall change points in the simulation scenarios.
Note that data from multiple risks provide information on the number and location of overall change points. On the other hand, a cause-specific change point is inferred from the event rate for a specific risk. This could be a potential explanation for why the mean squared error for $K$ is typically lower than for $K_r$ in
Web Figure~\ref{fig:simul_MSE_K}.

\subsection{Simulation study on restricting change point locations}
\label{sec:simul_restrict}

Section~\ref{sec:loc_cp} of the main manuscript discusses how the prior on the time locations of overall change points ${p(\bm\gamma\mid K)} = 1/\binom{|\mathcal{T}|}{K}$ is restricted to a subset of times $\mathcal{T}$.
Here, we empirically investigate the effect of such constraint.
Specifically, we compare with the alternative prior choice $p(\bm\gamma\mid K) = 1/\binom{t_{\max} - 1}{K}$ which does not have the restriction.
Firstly,
removing the restriction in the simulation study in Web Section~\ref{sec:simul_rep}
does not change results notably (results not shown):
there, most simulated data sets have observed events at all time points, such that the restriction does not change the prior ${p(\bm\gamma\mid K)}$ notably.

\begin{figure}
\centering
\includegraphics[width=\textwidth,page=1]{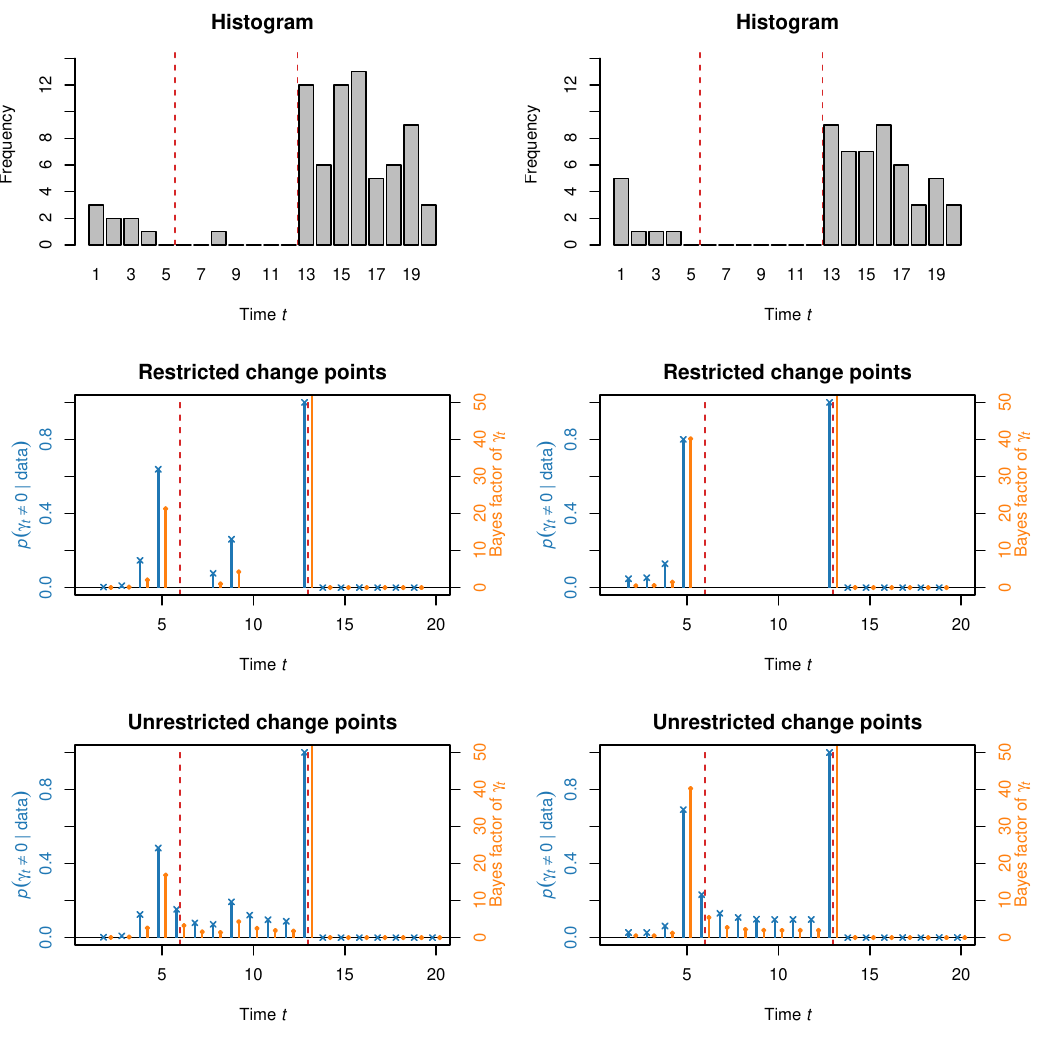}
\caption{Simulation study: Histograms of observed event times (top row), and
posterior probabilities ($\times$) and Bayes factors ($\bullet$)
for the presence of a change point for the model with restricted (middle row) and unrestricted (bottom row) change point locations
for the two simulated data sets (left and right columns) with no observed events near time $t=6$.
Orange lines correspond to Bayes factors, some of which are outside the plotting range.
The dashed red lines are drawn in correspondence of the true change points.}
\label{fig:simul_restrict_without}
\end{figure}

\begin{figure}
\centering
\includegraphics[width=\textwidth,page=2]{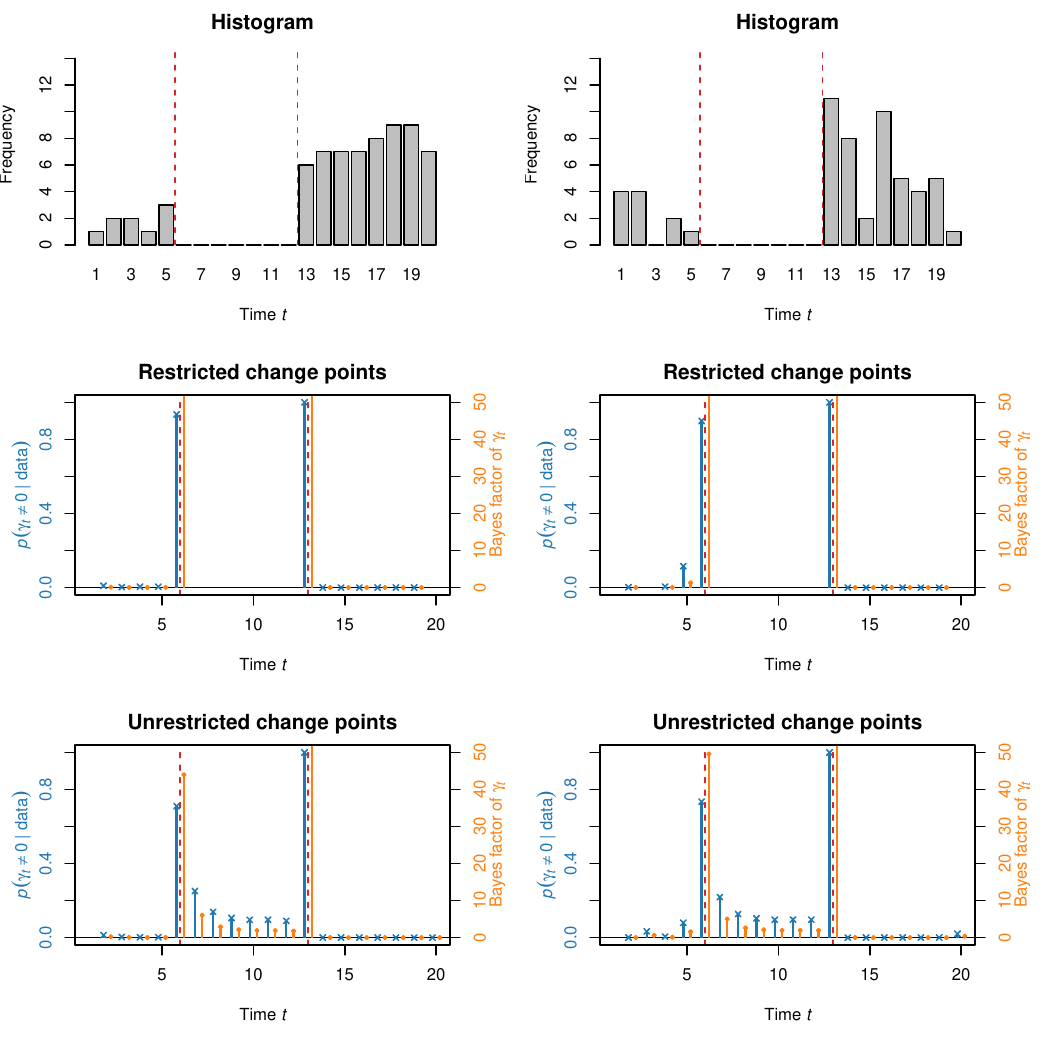}
\caption{Simulation study: Histograms of observed event times (top row), and
posterior probabilities ($\times$) and Bayes factors ($\bullet$)
for the presence of a change point for the model with restricted (middle row) and unrestricted (bottom row) change point locations
for the two simulated data sets (left and right columns) with observed events near time $t=6$.
Orange lines correspond to Bayes factors, some of which are outside the plotting range.
The dashed red lines are drawn in correspondence of the true change points.}
\label{fig:simul_restrict_with}
\end{figure}

We simulate data with $n=100$ observations, no predictors ($p=0$), $m=1$ risk, $t_{\max} = 15$, change points at $t=6,13$ and corresponding
baseline hazard across change points given by
$(\alpha_{1,1}^\star,\alpha_{1,2}^\star,\alpha_{1,3}^\star) = (-4,-9,-2)$.
We consider four replicated  data sets.
Two data sets, shown in Web Figure~\ref{fig:simul_restrict_without}, have no observed events near time $t=6$
and, thus, a change point at $t=6$ is not allowed in the Multivariate Bernoulli detector.
The other two, shown in Web Figure~\ref{fig:simul_restrict_with}, have observed events near $t=6$. Then, a change point at $t=6$ is allowed in the Multivariate Bernoulli detector.

For each replicate data set,
we fit the Multivariate Bernoulli detector as described in Section~\ref{sec:simul}
as well as the analogous model without a restriction on change point locations, i.e.\ with $p(\bm\gamma\mid K) = 1/\binom{t_{\max} - 1}{K}$.
The resulting inference on change point locations is summarized in Web Figures~\ref{fig:simul_restrict_without} and \ref{fig:simul_restrict_with}.
In Web Figure~\ref{fig:simul_restrict_without},
the restriction on change points does not allow inferring the true change point at $t=6$.
However, also without the restriction, the posterior probability of a change point at $t=5$ is higher than at $t=6$.
As such, the restriction does not reduce the quality of inference in the scenario where the true change point $t=5$ is not in the support of the restricted prior.
On the other hand, omitting the restriction results in posterior mass on spurious change points at times without observed event for all four replicate data sets (see both Web Figures~\ref{fig:simul_restrict_without} and \ref{fig:simul_restrict_with}).

\FloatBarrier
\section{Introduction to R package}
\label{ap:package}
% This appendix is based on R Markdown LaTeX output from the vignette of the R package mvb.detector.

The R package
\texttt{mvb.detector},
available from \url{https://github.com/willemvandenboom/mvb-detector},
implements the Multivariate Bernoulli detector. Here, we provide an introduction on how to use the
package.

\hypertarget{load-and-preprocess-data}{%
\subsection{Load and preprocess data}\label{load-and-preprocess-data}}

As example data, we consider the data set on unemployment duration from
\citet{McCall1996}, which is
available from the R package \texttt{Ecdat}.

\begin{Shaded}
\begin{Highlighting}[]
\FunctionTok{data}\NormalTok{(}\StringTok{"UnempDur"}\NormalTok{, }\AttributeTok{package =} \StringTok{"Ecdat"}\NormalTok{)}
\end{Highlighting}
\end{Shaded}

The data provide information on the time new employment after losing a
job. That is, the event of interest is finding a new job, and time to
event is measured from the loss of the previous job. The times are
recorded discretely, in two week intervals. For instance, \(T_i = 6\)
means that the time between jobs is twelve weeks. Furthermore, new
employment is either full-time or part-time, which represent \(m = 2\)
competing risks that can terminate the spell of unemployment.

For simplicity, we consider the subset of cases that are right-censored
(i.e. \texttt{censor4\ =\ 1}) or for which it is known whether the new
employment is full- or part-time (i.e.~\texttt{censor1\ =\ 1} or
\texttt{censor2\ =\ 1}).

\begin{Shaded}
\begin{Highlighting}[]
\NormalTok{UnempDur\_sub }\OtherTok{\textless{}{-}} \FunctionTok{subset}\NormalTok{(UnempDur, censor1 }\SpecialCharTok{|}\NormalTok{ censor2 }\SpecialCharTok{|}\NormalTok{ censor4)}
\end{Highlighting}
\end{Shaded}

Inspired by the function \texttt{dataLongCompRisks} from the R package
\texttt{discSurv}, \texttt{mvb.detector} requires that data are provided
as a data frame with the following variables:

\begin{enumerate}
\def\labelenumi{\arabic{enumi}.}
\tightlist
\item
  As first variable, an integer vector \texttt{time} with the
  time-to-event data
\item
  As second variable, a factor \texttt{event} of event types/causes with
  the first level indicating that the time was censored
\item
  Optionally, additional variables that are treated as predictor values
\end{enumerate}

We create such data frame where we include \texttt{ui} and
standardized \texttt{disrate} as covariates:

\begin{enumerate}
\def\labelenumi{\arabic{enumi}.}
\tightlist
\item
  The variable \texttt{ui} is a binary indicator of whether the person received
  benefits from an unemployment insurance scheme.
\item
  The variable \texttt{disrate} is the disregard rate: the \emph{disregard} is the
  amount that a person is allowed to earn in a new job without reduction
  in unemployment benefits. The \emph{disregard rate} is the disregard
  divided by the earnings in the lost job.
\end{enumerate}

Finally, we right-censor all unemployment spells longer than \(t=10\) to
reduce the range of time points for ease of exposition.

\begin{Shaded}
\begin{Highlighting}[]
\NormalTok{UnempDur\_sub}\SpecialCharTok{$}\NormalTok{event }\OtherTok{\textless{}{-}}\NormalTok{ 0L}
\NormalTok{UnempDur\_sub}\SpecialCharTok{$}\NormalTok{event[UnempDur\_sub}\SpecialCharTok{$}\NormalTok{censor1 }\SpecialCharTok{==} \DecValTok{1}\NormalTok{] }\OtherTok{\textless{}{-}}\NormalTok{ 1L}
\NormalTok{UnempDur\_sub}\SpecialCharTok{$}\NormalTok{event[UnempDur\_sub}\SpecialCharTok{$}\NormalTok{censor2 }\SpecialCharTok{==} \DecValTok{1}\NormalTok{] }\OtherTok{\textless{}{-}}\NormalTok{ 2L}

\CommentTok{\# Right{-}censor unemployment spells longer than 10.}
\NormalTok{UnempDur\_sub}\SpecialCharTok{$}\NormalTok{event[UnempDur\_sub}\SpecialCharTok{$}\NormalTok{spell }\SpecialCharTok{\textgreater{}} \DecValTok{10}\NormalTok{] }\OtherTok{\textless{}{-}}\NormalTok{ 0L}
\NormalTok{UnempDur\_sub}\SpecialCharTok{$}\NormalTok{spell }\OtherTok{\textless{}{-}} \FunctionTok{pmin}\NormalTok{(}\DecValTok{10}\NormalTok{, UnempDur\_sub}\SpecialCharTok{$}\NormalTok{spell)}

\NormalTok{UnempDur\_sub}\SpecialCharTok{$}\NormalTok{event }\OtherTok{\textless{}{-}} \FunctionTok{factor}\NormalTok{(}\AttributeTok{x =}\NormalTok{ UnempDur\_sub}\SpecialCharTok{$}\NormalTok{event, }\AttributeTok{levels =} \DecValTok{0}\SpecialCharTok{:}\DecValTok{2}\NormalTok{)}
\FunctionTok{levels}\NormalTok{(UnempDur\_sub}\SpecialCharTok{$}\NormalTok{event) }\OtherTok{\textless{}{-}} \FunctionTok{c}\NormalTok{(}\StringTok{"censored"}\NormalTok{, }\StringTok{"Full{-}time"}\NormalTok{, }\StringTok{"Part{-}time"}\NormalTok{)}

\NormalTok{data }\OtherTok{\textless{}{-}} \FunctionTok{data.frame}\NormalTok{(}
  \AttributeTok{time =} \FunctionTok{as.integer}\NormalTok{(UnempDur\_sub}\SpecialCharTok{$}\NormalTok{spell), }\AttributeTok{event =}\NormalTok{ UnempDur\_sub}\SpecialCharTok{$}\NormalTok{event,}
  \AttributeTok{X =} \FunctionTok{cbind}\NormalTok{(UnempDur\_sub}\SpecialCharTok{$}\NormalTok{ui }\SpecialCharTok{==} \StringTok{"yes"}\NormalTok{, }\FunctionTok{scale}\NormalTok{(UnempDur\_sub}\SpecialCharTok{$}\NormalTok{disrate))}
\NormalTok{)}

\FunctionTok{summary}\NormalTok{(data)}
\end{Highlighting}
\end{Shaded}

\begin{verbatim}
##       time              event           X.1              X.2         
##  Min.   : 1.000   censored :1449   Min.   :0.0000   Min.   :-1.4689  
##  1st Qu.: 2.000   Full-time: 919   1st Qu.:0.0000   1st Qu.:-0.7870  
##  Median : 5.000   Part-time: 299   Median :1.0000   Median :-0.1050  
##  Mean   : 5.311                    Mean   :0.5778   Mean   : 0.0000  
##  3rd Qu.: 9.000                    3rd Qu.:1.0000   3rd Qu.: 0.5352  
##  Max.   :10.000                    Max.   :1.0000   Max.   :12.6986
\end{verbatim}

\hypertarget{fit-the-multivariate-bernoulli-detector}{%
\subsection{Fit the Multivariate Bernoulli
detector}\label{fit-the-multivariate-bernoulli-detector}}

We fit the Multivariate Bernoulli detector to the data using 20000
Markov chain Monte Carlo (MCMC) iterations, discarding the first 10000
as burn-in.

\begin{Shaded}
\begin{Highlighting}[]
\FunctionTok{set.seed}\NormalTok{(1L)  }\CommentTok{\# Set seed for reproducibility.}

\NormalTok{mvbd\_fit }\OtherTok{\textless{}{-}}\NormalTok{ mvb.detector}\SpecialCharTok{::}\FunctionTok{run\_mvb\_detector}\NormalTok{(}
  \AttributeTok{data =}\NormalTok{ data, }\AttributeTok{n\_iter =} \FloatTok{2e4}\NormalTok{L, }\AttributeTok{burnin =} \FloatTok{1e4}\NormalTok{L}
\NormalTok{)}
\end{Highlighting}
\end{Shaded}

\hypertarget{summarize-posterior-inference}{%
\subsection{Summarize posterior
inference}\label{summarize-posterior-inference}}

The MCMC iterations are now saved in \texttt{mvbd\_fit}. The package
provides a variety of functions to summarize the corresponding posterior
inference. Firstly, we plot the posterior on the number of change
points, and find that the posterior concentrates on two change points.

\begin{Shaded}
\begin{Highlighting}[]
\NormalTok{mvb.detector}\SpecialCharTok{::}\FunctionTok{plot\_K}\NormalTok{(mvbd\_fit)}
\end{Highlighting}
\end{Shaded}

\includegraphics[width=3in,height=3in]{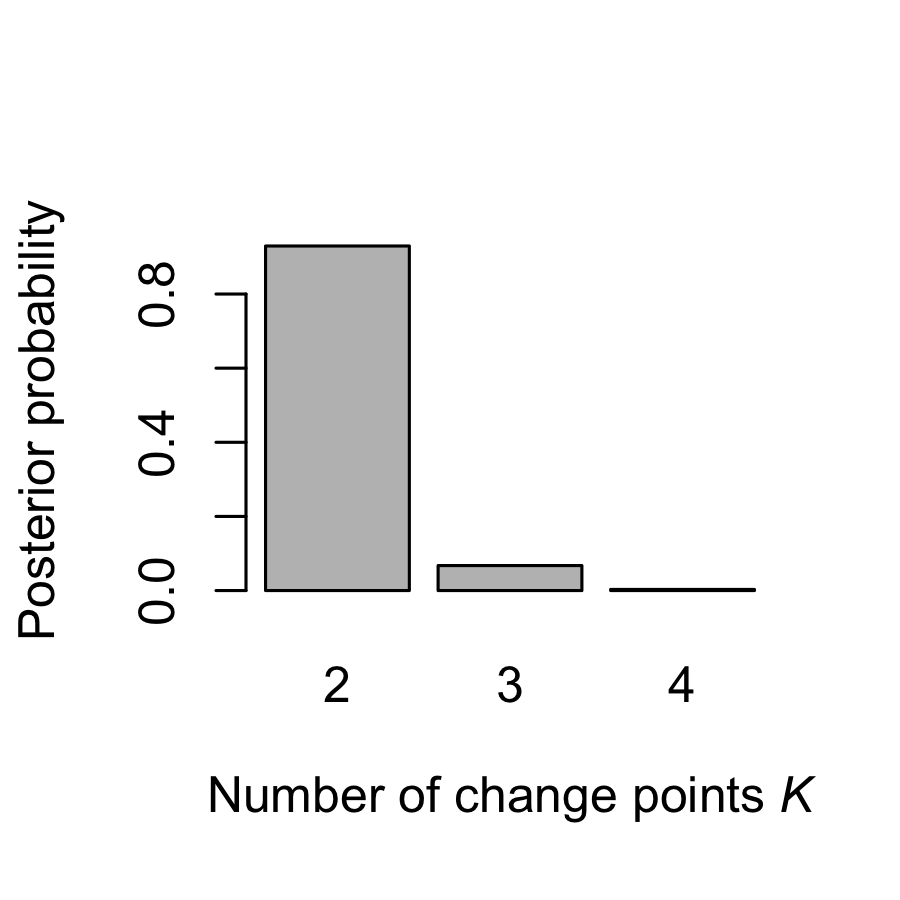}

To inspect the location of the change points, we plot posterior
inclusion probabilities and Bayes factors:

\begin{Shaded}
\begin{Highlighting}[]
\FunctionTok{par}\NormalTok{(}\AttributeTok{mar =} \FunctionTok{c}\NormalTok{(}\DecValTok{5}\NormalTok{, }\DecValTok{4}\NormalTok{, }\DecValTok{4}\NormalTok{, }\DecValTok{4}\NormalTok{))  }\CommentTok{\# Increase right margin to make space for labels.}
\NormalTok{mvb.detector}\SpecialCharTok{::}\FunctionTok{plot\_change\_points}\NormalTok{(mvbd\_fit)}
\end{Highlighting}
\end{Shaded}

\includegraphics[width=3in,height=3in]{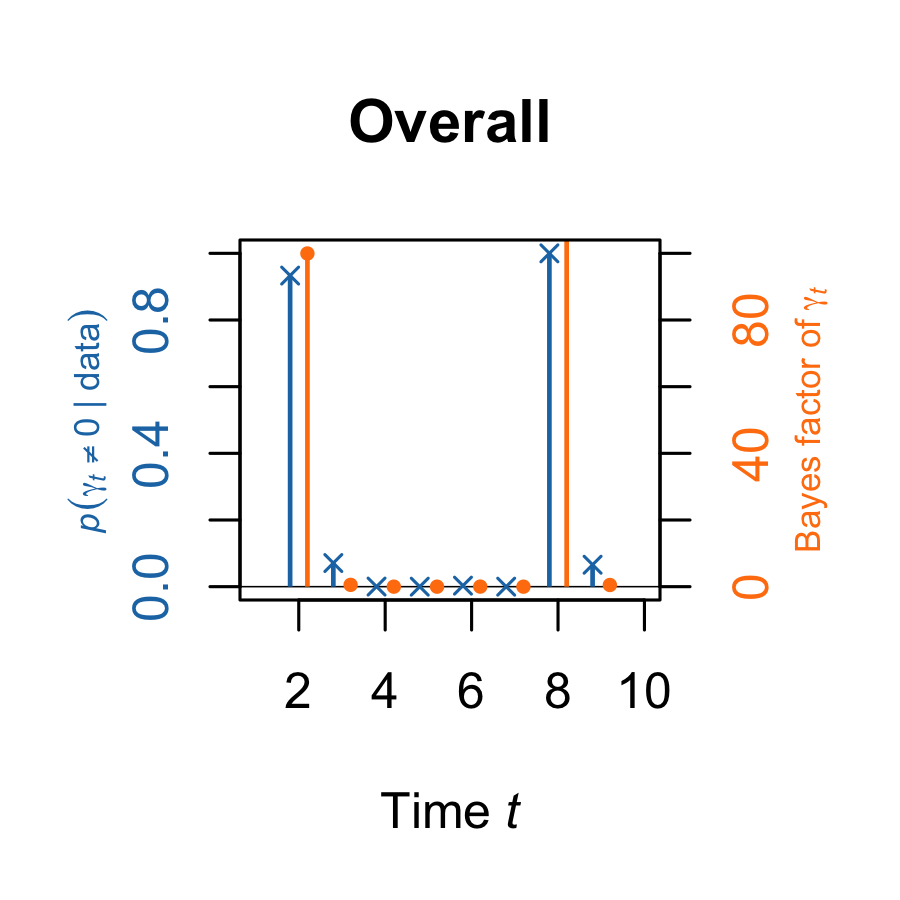}
\includegraphics[width=3in,height=3in]{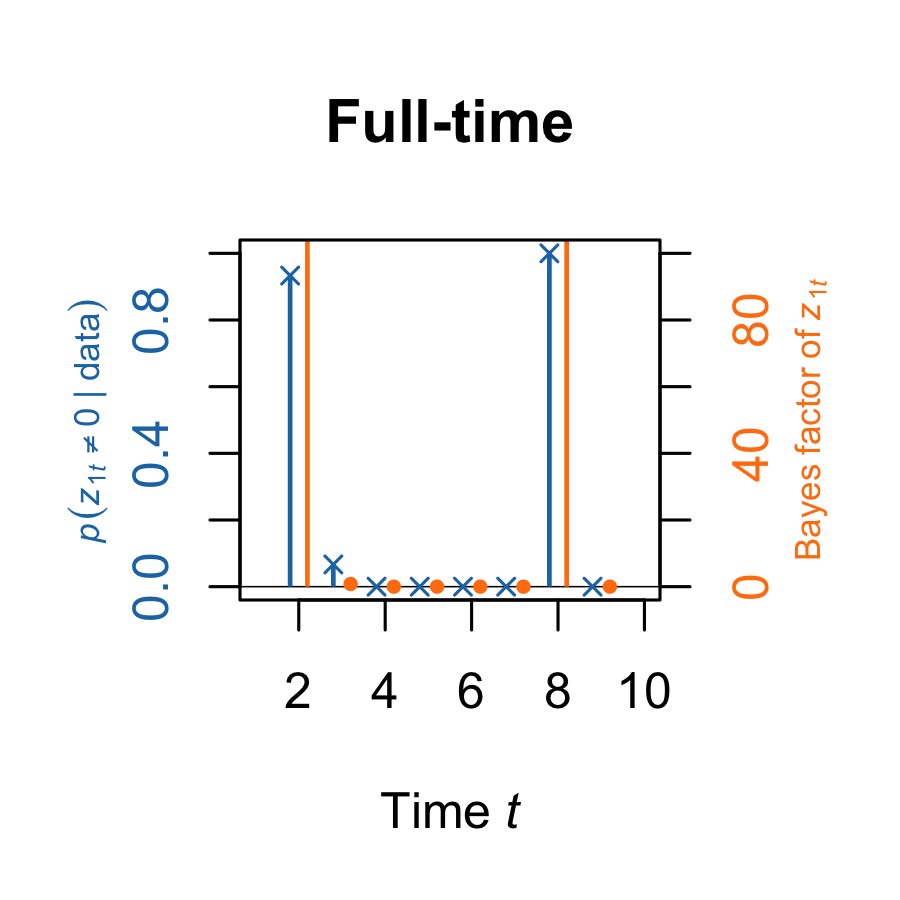}
\includegraphics[width=3in,height=3in]{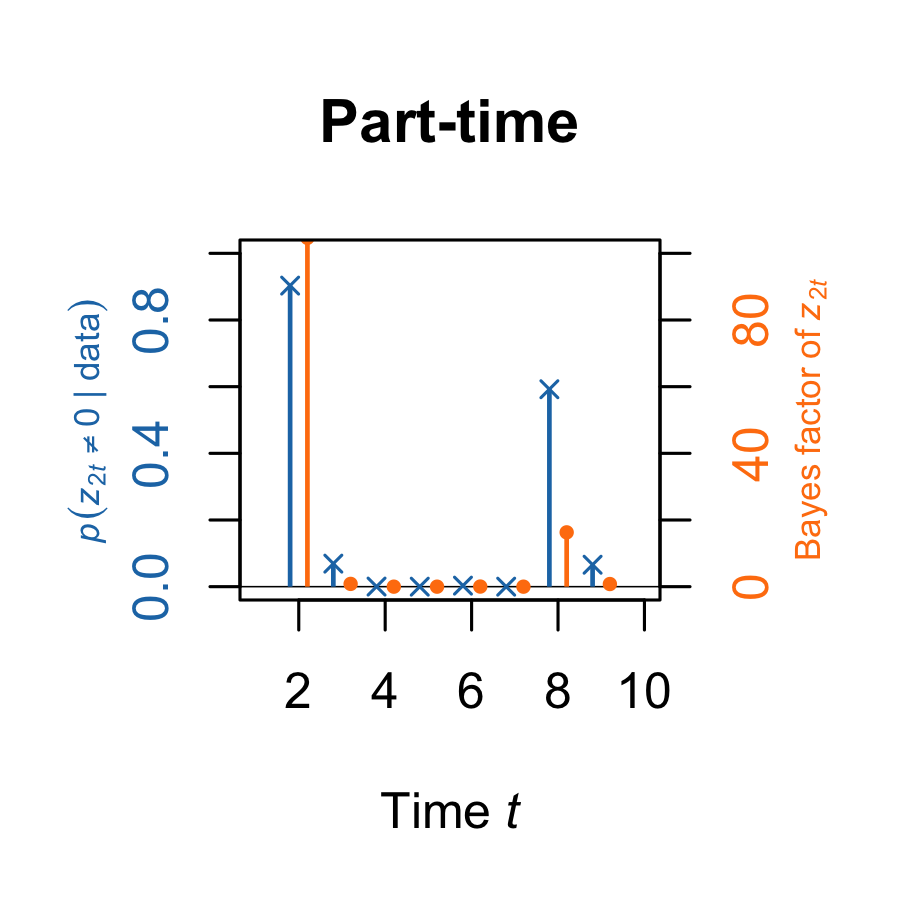}

The change points are mostly at \(t=2\) and \(t=8\) in the posterior
distribution. The second change point (\(t=8\)) might not be present for
the hazard rate specific to part-time re-employment. The posterior mean
and 95\% credible intervals of the baseline hazards are consistent with
this:

\begin{Shaded}
\begin{Highlighting}[]
\NormalTok{mvb.detector}\SpecialCharTok{::}\FunctionTok{plot\_baseline\_hazards}\NormalTok{(mvbd\_fit)}
\end{Highlighting}
\end{Shaded}

\includegraphics[width=3in,height=3in]{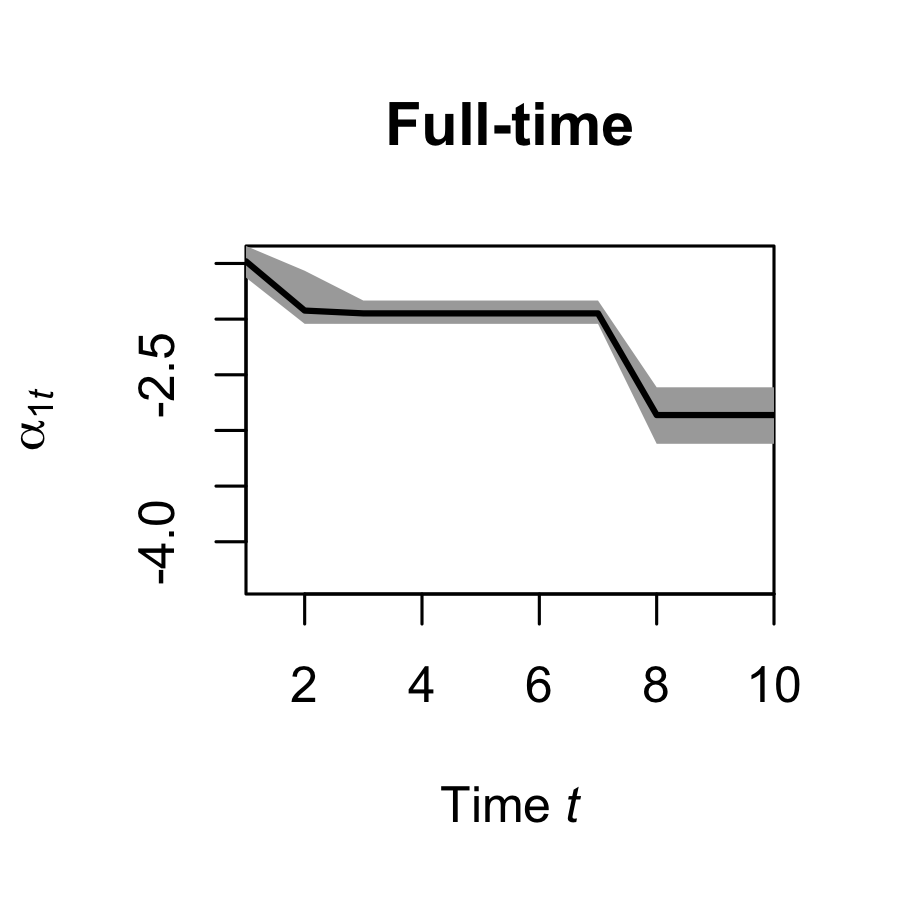}
\includegraphics[width=3in,height=3in]{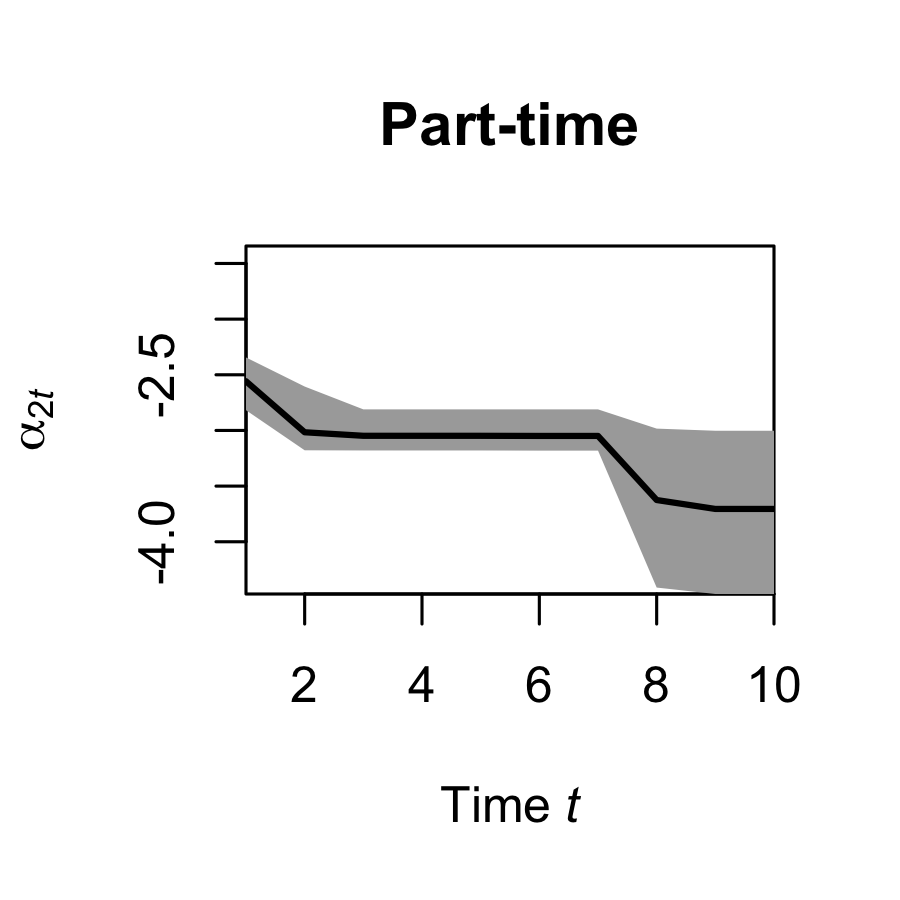}

For the regression coefficients, we can compute posterior inclusion
probabilities, and plot posterior means and 95\% credible intervals. In
the output below, no effect of disregard rate on part-time re-employment
is inferred. At the same time, the other covariate effects are negative:
employment benefits are associated with a longer time to re-employment.
The same holds for disregard rate and full-time re-employment. A potential explanation is that
those with a higher disregard rate prefer to take a part-time job
instead of a full-time job to maximize unemployment benefits, which is
in line with Table~III of \citet{McCall1996}.

\begin{Shaded}
\begin{Highlighting}[]
\NormalTok{pred\_names }\OtherTok{\textless{}{-}} \FunctionTok{c}\NormalTok{(}\StringTok{"Unemployment insurance"}\NormalTok{, }\StringTok{"Disregard rate"}\NormalTok{)}

\NormalTok{mvb.detector}\SpecialCharTok{::}\FunctionTok{compute\_post\_inc\_prob}\NormalTok{(}
  \AttributeTok{mvbd\_fit =}\NormalTok{ mvbd\_fit, }\AttributeTok{pred\_names =}\NormalTok{ pred\_names, }\AttributeTok{digits =}\NormalTok{ 3L}
\NormalTok{)}
\end{Highlighting}
\end{Shaded}

\begin{verbatim}
##                        Full-time Part-time
## Unemployment insurance         1     1.000
## Disregard rate                 1     0.001
\end{verbatim}

\begin{Shaded}
\begin{Highlighting}[]
\FunctionTok{par}\NormalTok{(}\AttributeTok{mar =} \FunctionTok{c}\NormalTok{(}\DecValTok{5}\NormalTok{, }\DecValTok{10}\NormalTok{, }\DecValTok{4}\NormalTok{, }\DecValTok{2}\NormalTok{))  }\CommentTok{\# Increase left margin to makes space for labels.}
\NormalTok{mvb.detector}\SpecialCharTok{::}\FunctionTok{plot\_regression}\NormalTok{(}\AttributeTok{mvbd\_fit =}\NormalTok{ mvbd\_fit, }\AttributeTok{pred\_names =}\NormalTok{ pred\_names)}
\end{Highlighting}
\end{Shaded}

\includegraphics[width=5in]{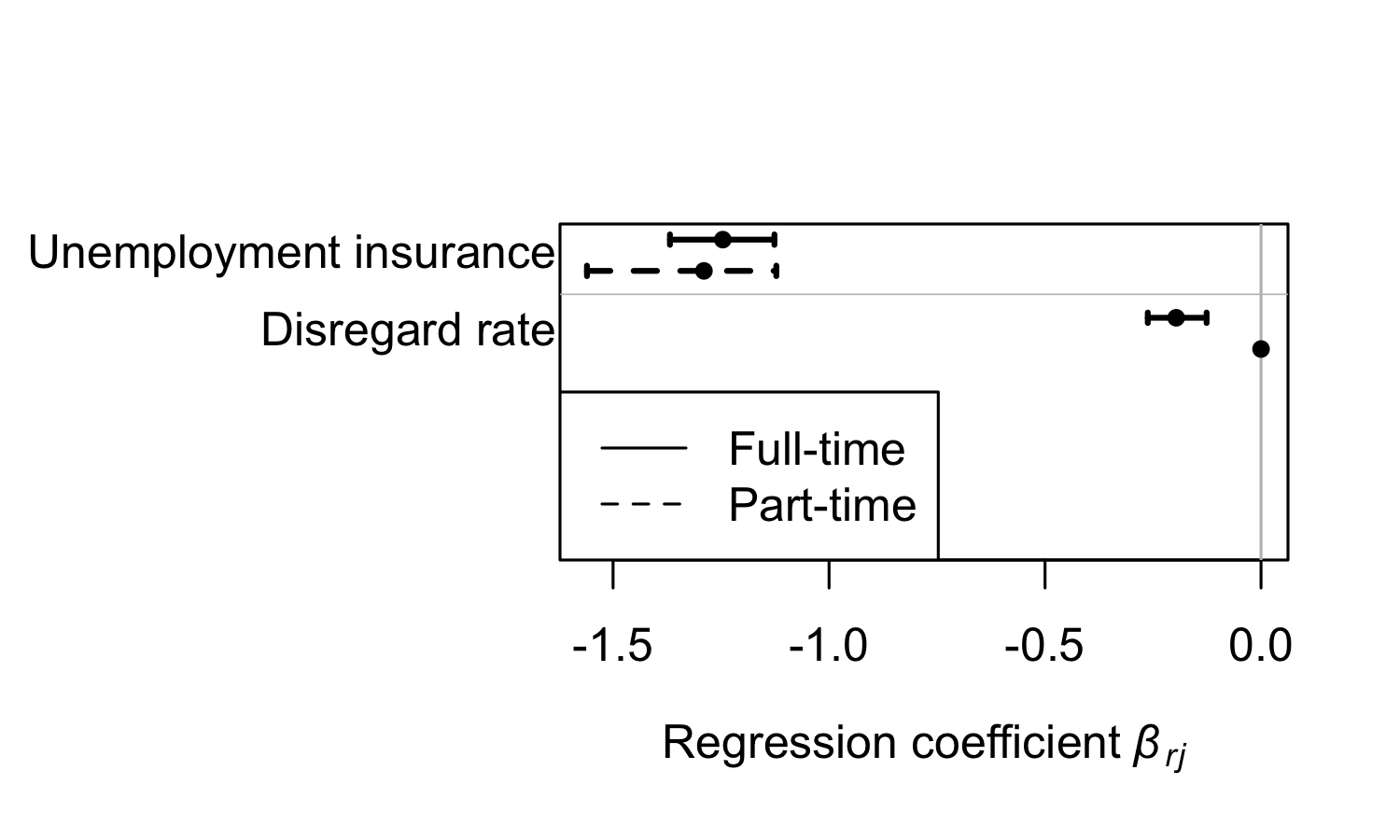}

\FloatBarrier
\pagebreak
\section{Additional figures}
\label{ap:fig}

\begin{figure}[h]
\centering
\includegraphics[width=\textwidth]{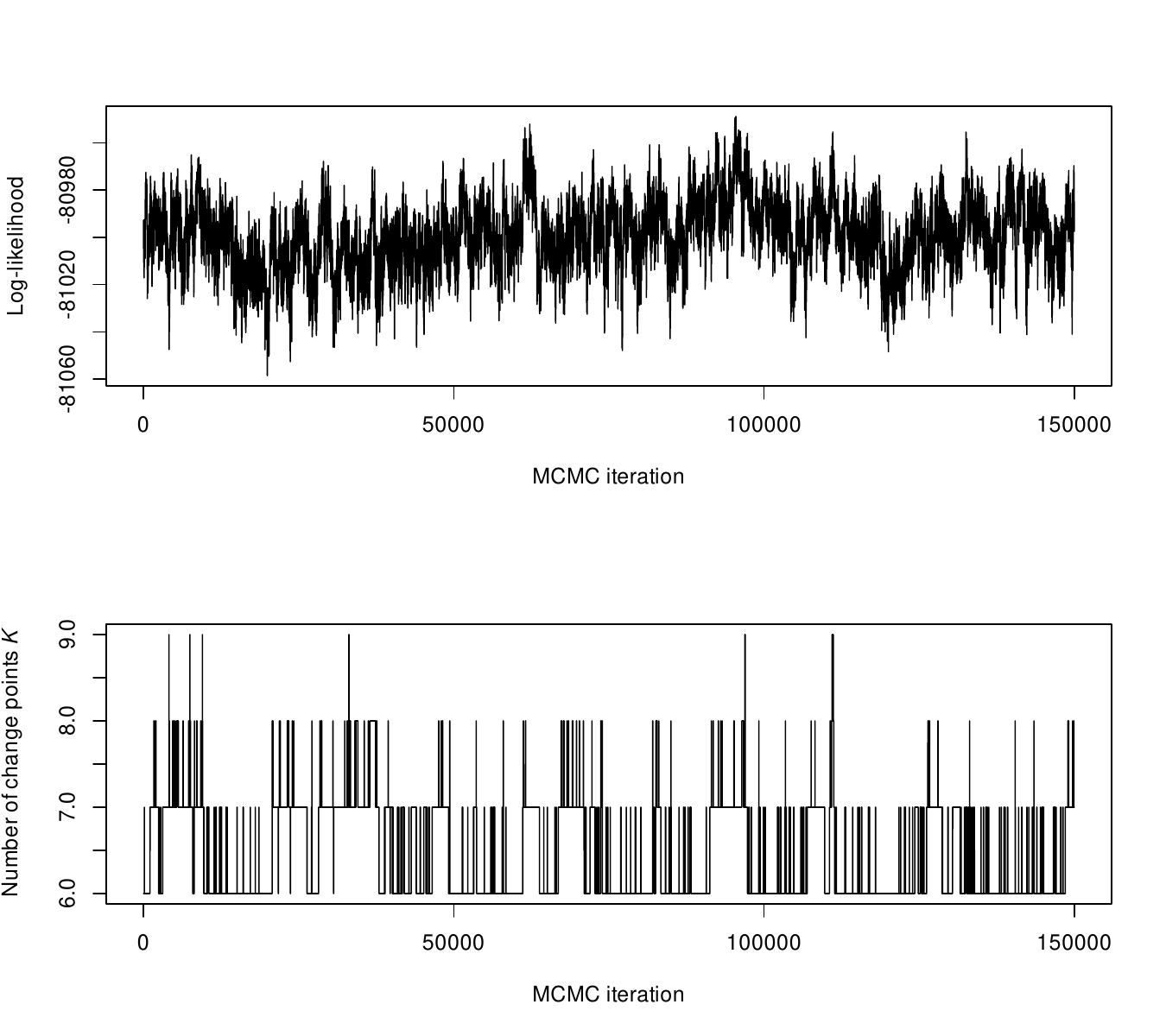}
\caption{ICU data: Trace plots of the MCMC chain for the log-likelihood $\log\{p({\textnormal{data}\mid\bm\theta})\}$ and number of change points $K$.
\label{fig:mimic_mcmc}}
\end{figure}

\begin{figure}
\centering
\includegraphics[width=\textwidth]{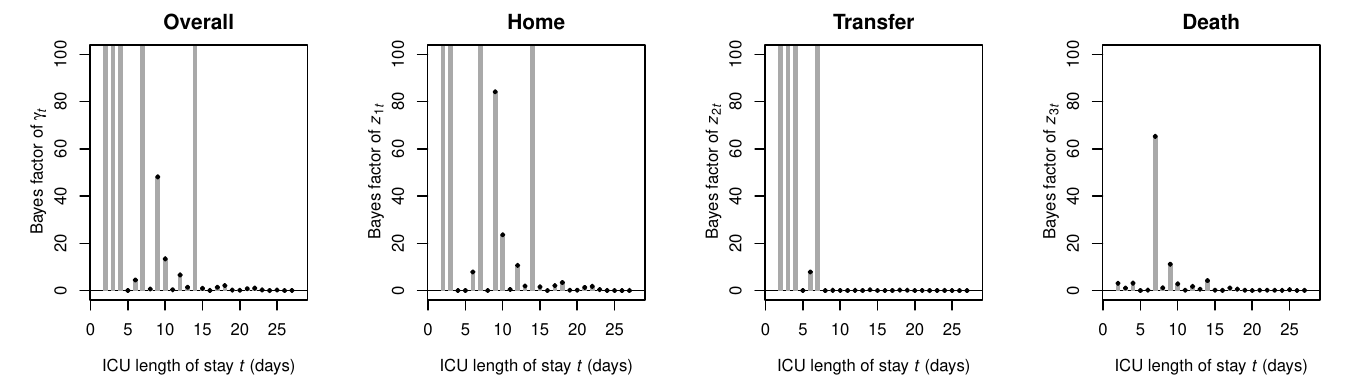}
\caption{ICU data: Bayes factors for the presence of a change point for the overall (left column) and cause-specific (other columns) hazard functions.
The gray lines correspond to Bayes factors, some of which are outside the plotting range.
\label{fig:mimic_hazard_BF}}
\end{figure}

\begin{figure}
\centering
\includegraphics[width=\textwidth]{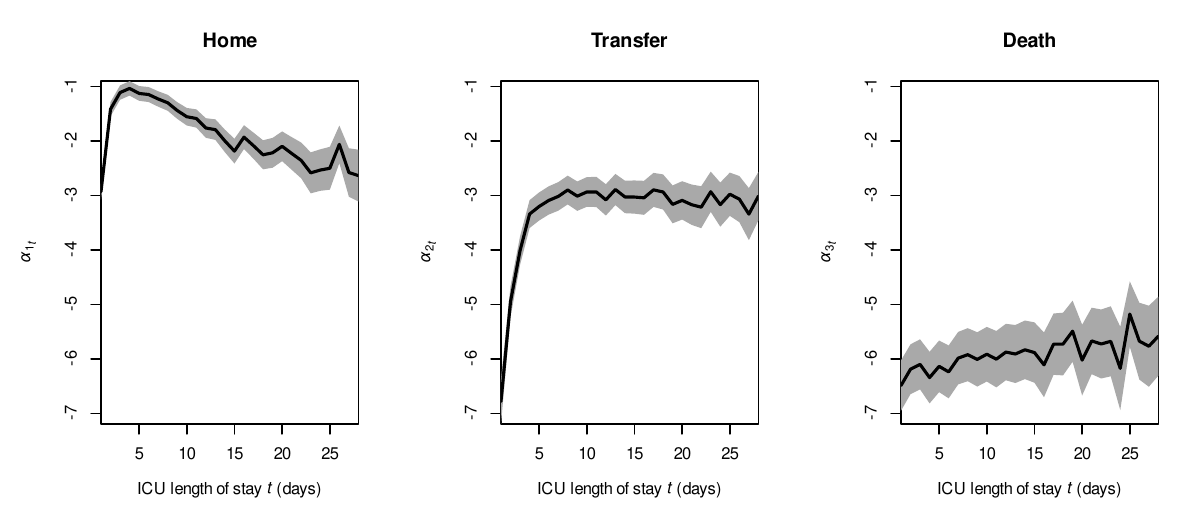}
\caption{ICU data: Maximum likelihood estimates of the baseline hazard parameter $\alpha_{rt}$ (lines) with their 95\% confidence intervals demarcated by shaded areas.}
\label{fig:mimic_nnet_hazard}
\end{figure}

\begin{figure}
\centering
\includegraphics[width=0.5\textwidth]{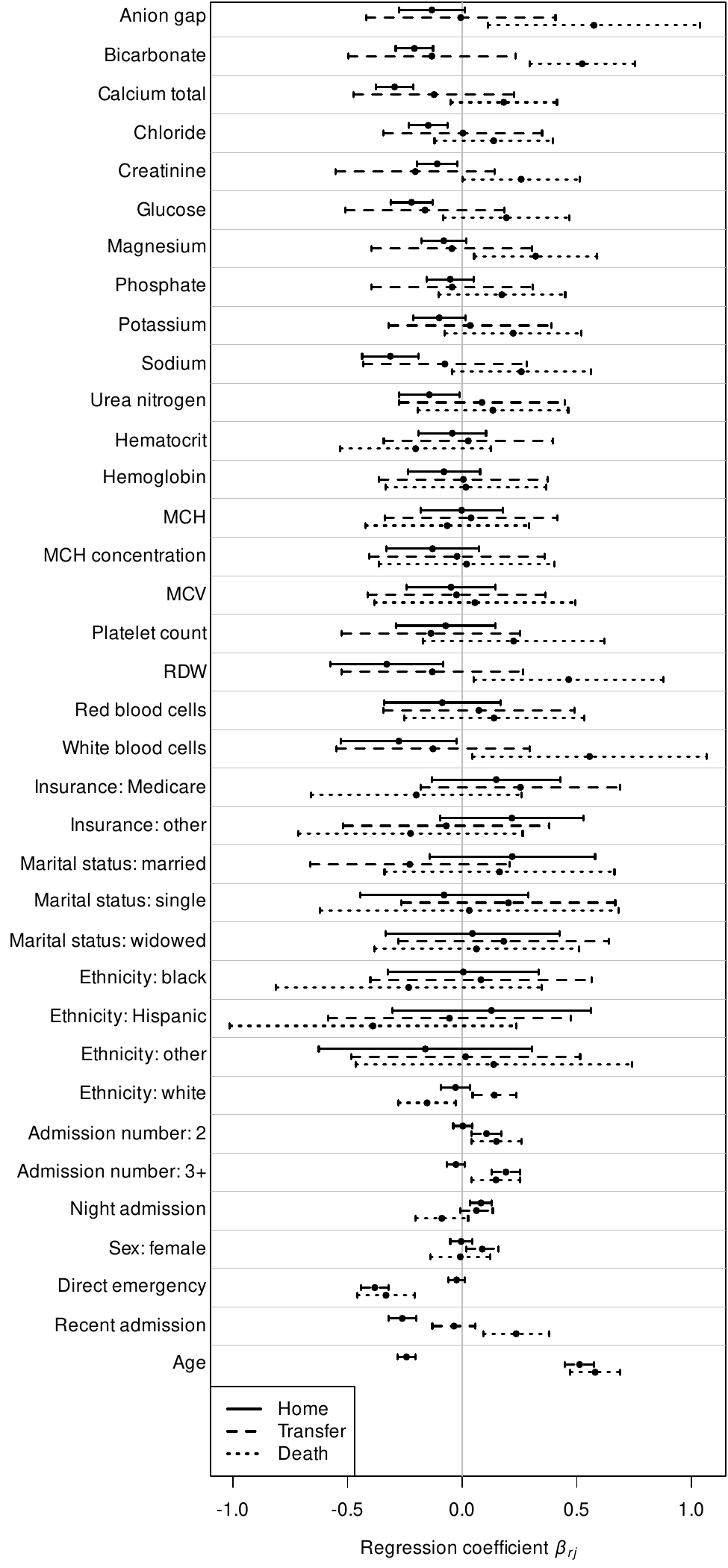}
\caption{ICU data: Maximum likelihood estimates (dot) and 95\% confidence intervals (lines) of the regression coefficients for each risk. The categorical predictors are coded as dummy variables as detailed in Web Appendix~\ref{ap:mimic}. MCH stands for mean cell hemoglobin, MCV for mean corpuscular volume and RDW for red blood cell distribution width.
}
\label{fig:mimic_nnet_regression}
\end{figure}

\begin{figure}
\centering
\includegraphics[width=\textwidth]{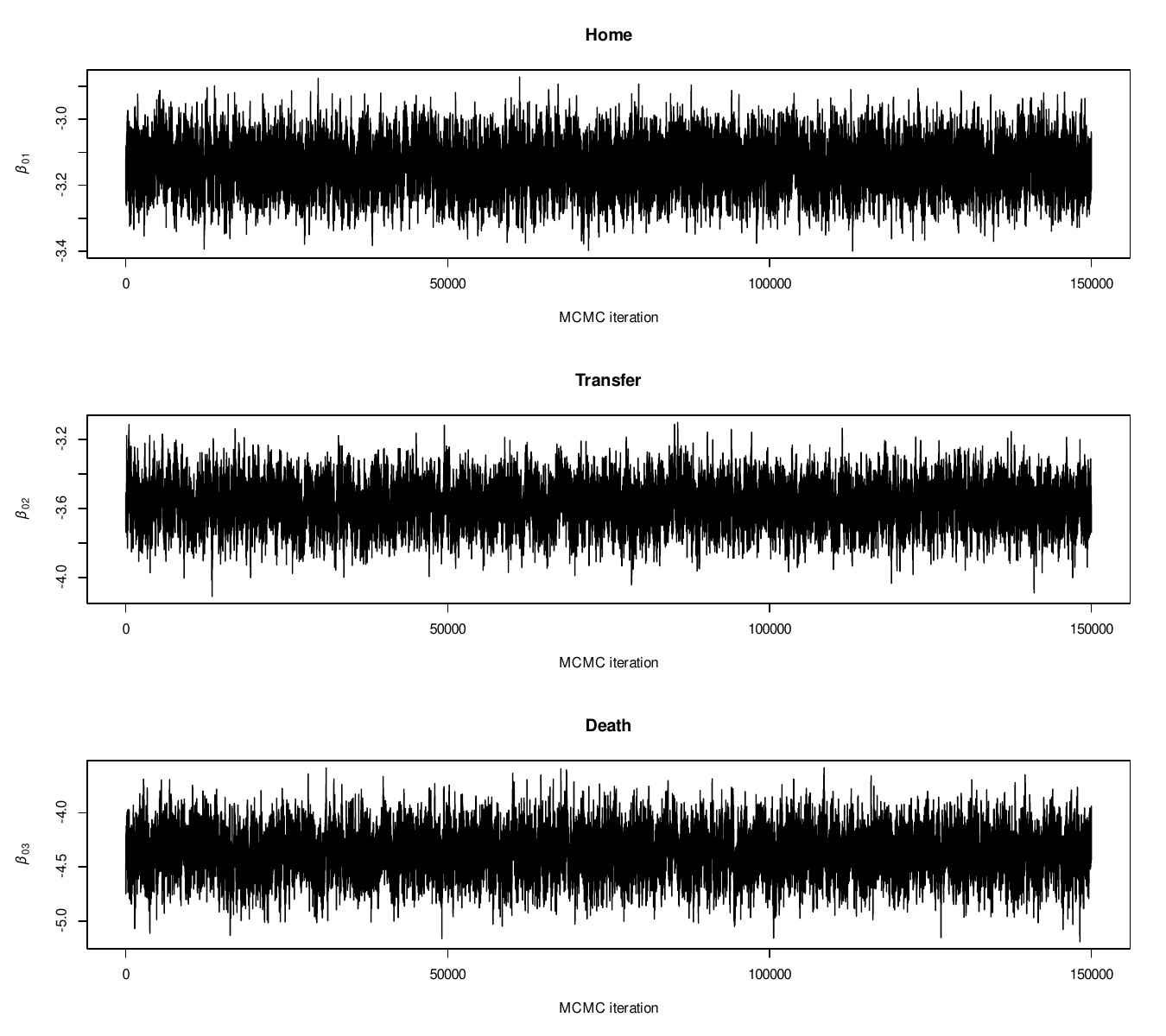}
\caption{ICU data: Trace plots of the MCMC chain for the intercepts $\beta_{0r}$ in the model by \citet{King2021}. Trace plots for other parameters show similarly good mixing (not shown).
\label{fig:brea_mcmc}}
\end{figure}

\begin{figure}
\centering
\includegraphics[width=\textwidth]{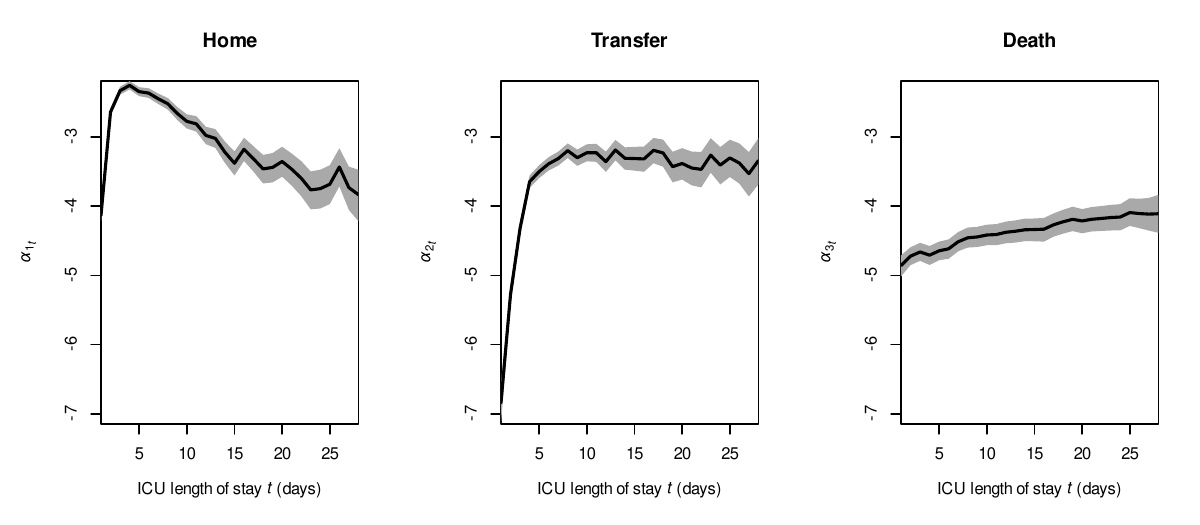}
\caption{ICU data: Posterior inference on the baseline hazard parameter $\alpha_{rt}$ from the model by \citet{King2021}. Lines represent posterior means and shaded areas correspond to 95\% credible intervals.}
\label{fig:mimic_brea_hazard}
\end{figure}

\begin{figure}
\centering
\includegraphics[width=\textwidth]{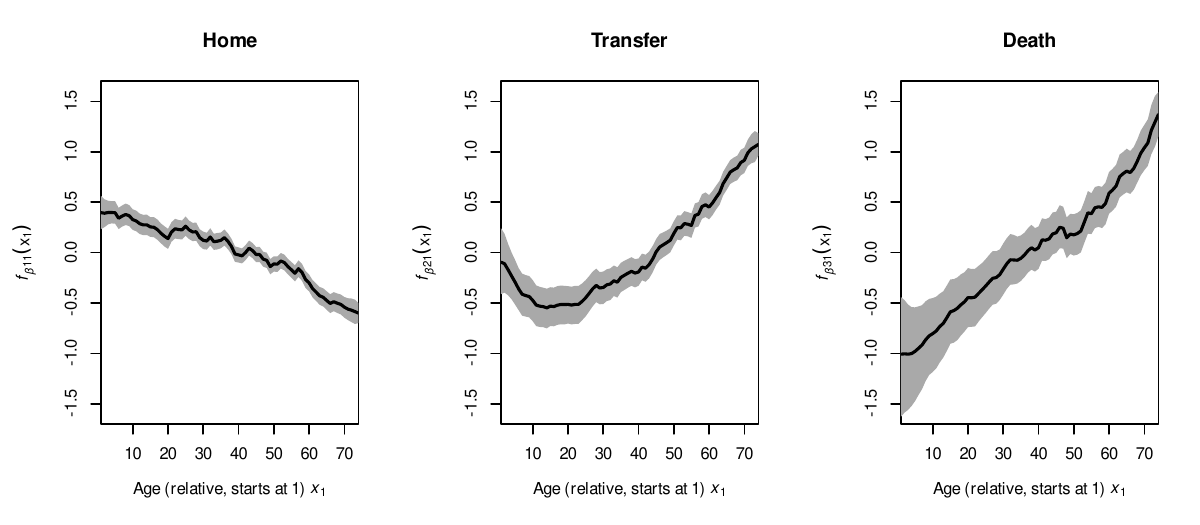}
\caption{ICU data: Posterior inference on $f_{\beta r 1}(x_1)$, where $x_1$ denotes \emph{age}, obtained with 
the model by \citet{King2021}. Lines represent posterior means and shaded areas correspond to 95\% credible intervals. Here, age is in years on a relative scale with $x_1 = 1$ corresponding to the youngest patient and $x_1=74$ with the oldest.}
\label{fig:mimic_brea_regression_age}
\end{figure}

\begin{figure}
\centering
\includegraphics[width=0.5\textwidth]{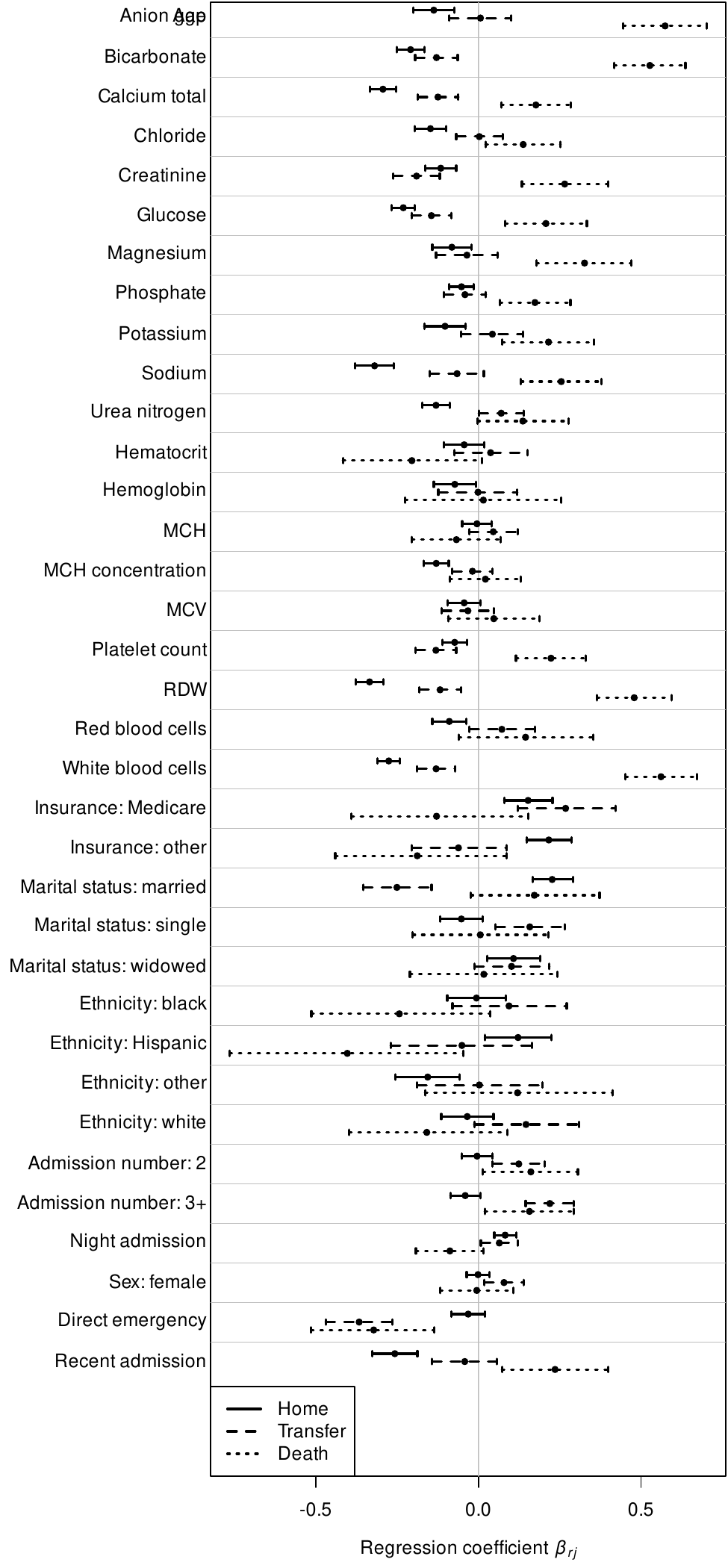}
\caption{ICU data: Maximum likelihood estimates (dot) and 95\% confidence intervals (lines) of the regression coefficients for the binary predictors from the model by \citet{King2021} for each risk. The categorical covariates are coded as dummy variables as detailed in Web Appendix~\ref{ap:mimic}. MCH stands for mean cell hemoglobin, MCV for mean corpuscular volume and RDW for red blood cell distribution width.
}
\label{fig:mimic_brea_regression}
\end{figure}

%% Bibliography, a section that is not referred to "Web Appendix".
\makeatletter
\def\section{\@startsection{section}{1}{\z@ }%
  {-3.5ex\@plus -1ex\@minus -.2ex}{2.3ex \@plus .2ex}%
  {\noindent\normalfont \Large \bfseries}%
}
\makeatother

\FloatBarrier
\bibliographystyle{biom}
\bibliography{Ref_discretetimesurvival}